\documentclass[aps,prx,twocolumn,superscriptaddress,amsmath,nofootinbib,amssymb,10pt,floatfix]{revtex4-1}
\usepackage{graphicx}
\usepackage{ulem}   
\usepackage{xcolor}
\usepackage{bbold}
\usepackage{comment}
\usepackage[colorlinks, linkcolor=blue, citecolor=blue]{hyperref}
\newcommand{\be}{\begin{equation}}
\newcommand{\ee}{\end{equation}}
\newcommand{\beq}{\begin{eqnarray}}
\newcommand{\eeq}{\end{eqnarray}}
\newcommand{\ba}{\[\begin{aligned}}
\newcommand{\ea}{\end{aligned}\]}

\newcommand{\la}{\langle}
\newcommand{\ra}{\rangle}

\renewcommand{\vec}[1]{\boldsymbol{#1}}
\renewcommand{\hat}[1]{{\bf {\widehat #1}}}
\renewcommand{\phi}{\varphi}
\renewcommand{\epsilon}{\varepsilon}

\newcommand{\bb}[1]{\mathbb{#1}}

\def \q {\vec{q}}

\def \O {{\cal{O}}}
\def \N {{\cal{N}}}

\def \r {\vec{r}}
\def \ve{{\varepsilon}}

\def \L{{\cal{L}}}
\def \S{{\cal{S}}}
\def \a{{\vec{a}}}

\def \k{{\vec{k}}}
\def \Q{{\vec{Q}}}

\def \ra{{\rangle}}
\def \la{{\langle}}
\def \tn{\textnormal}

\def \ba{\begin{align*}}
\def \ea{\end{align*}}

\newcounter{indice}

\def \bs{\boldsymbol}

\begin{document}
\title{Theory of a continuous bandwidth-tuned Wigner-Mott transition}

\author{Seth Musser}
\author{T. Senthil}
\affiliation{Department of Physics, Massachusetts Institute of Technology, Cambridge MA 02139}
\author{Debanjan Chowdhury}
\affiliation{Department of Physics, Cornell University, Ithaca NY 14853}
\begin{abstract}
We develop a theory for a continuous bandwidth-tuned transition at fixed \textit{fractional} electron filling from a metal with a generic Fermi surface to a `Wigner-Mott' insulator that spontaneously breaks crystalline space-group symmetries. Across the quantum critical point, (i) the entire electronic Fermi surface disappears abruptly upon approaching from the metallic side, and (ii) the insulating charge gap and various order parameters associated with the spontaneously broken space-group symmetries vanish continuously upon approaching from the insulating side. Additionally, the insulating side hosts a Fermi surface of neutral spinons. We present a framework for describing such continuous metal-insulator transitions (MITs) and analyze the example of a bandwidth-tuned transition at a filling, $\nu=1/6$, for spinful electrons on the triangular lattice. By extending the theory to a certain large-$N$ limit, we provide a concrete example of such a continuous MIT and discuss numerous experimental signatures near the critical point. We place our results in the context of recent experiments in moir\'e transition metal dichalcogenide materials.
\end{abstract}

\maketitle

\section{Introduction}
\label{sec:intro}

Conventional quantum phase transitions (QPTs) in insulators associated with the onset of spontaneously broken symmetries can be described using the classic Landau-Ginzburg-Wilson framework. The critical field theory is governed by the long-wavelength and low-energy fluctuations of a local `order parameter’. Continuous QPTs in metals are significantly more challenging to describe theoretically due to the abundance of low-energy gapless excitations in the vicinity of an electronic Fermi surface (FS). Arguably the most intriguing example \cite{senthil2008critical}  of such a {\it continuous} QPT in a metal is associated with the {\it abrupt} disappearance of an entire electronic FS, as the metal evolves into an electrical insulator at a fixed density. A theory for such  a continuous Mott transition at half filling $\nu_c = 1/2$ was described in Ref. \onlinecite{senthil_theory_2008}. The corresponding Mott insulating state is a quantum spin liquid that preserves all symmetries of the underlying microscopic Hamiltonian.  In this article, we focus on a particular class of such continuous bandwidth-tuned metal-insulator transitions (MITs) at fixed electron filling, $\nu_c < 1/2$, where the insulator spontaneously breaks underlying crystalline space-group symmetries. Such a state is often  dubbed  a ``Wigner-Mott"  (WM) insulator, and we will use that terminology.  The evolution from a Wigner-Mott insulator to a symmetry preserving Fermi liquid metal raises a number of fascinating and deep theoretical questions, and further may be experimentally accessible in the near future.

There have been a number of recent theoretical and experimental breakthroughs in realizing interaction-induced insulators at partial filling of moir\'e flat bands in bilayers of transition metal dichalcogenide (TMD) materials \cite{padhi_generalized_2021}. These include the observation of a robust Mott insulator at half filling of the moir\'e unit cell \cite{tang_simulation_2020} accompanied by the formation of local moments and a panoply of Wigner-Mott insulators at a sequence of other commensurate fillings \cite{regan_mott_2020, xu_correlated_2020, huang_correlated_2021}. The WM insulators are expected to display a variety of translational and/or rotational symmetry breaking, some of which have been observed experimentally. More recently, two independent works have provided compelling evidence for a continuous bandwidth-tuned transition from a Mott insulator to a Fermi liquid metal at fixed $\nu_c=1/2$ \cite{li_continuous_2021, ghiotto_quantum_2021}. In addition to a scaling collapse of the electrical resistivity across the MIT, the experiment finds the charge gap and the inverse Fermi velocity to vanish continuously upon approaching the transition from the insulating and metallic sides, respectively. Magnetic measurements reveal a smooth evolution of the susceptibility across the MIT and no sign of any magnetic ordering in the Mott insulator down to the lowest temperatures. Numerous aspects of the observed phenomenology are reminiscent of a continuous transition \cite{senthil_theory_2008} from a Fermi liquid metal to a paramagnetic Mott insulator with a spinon Fermi surface. Additionally an exact diagonalization approach lends support to a continuous MIT at $\nu_c=1/2$ in the TMD setting, or at least a weakly first-order transition \cite{morales-duran_metal-insulator_2021}.

\begin{figure}
    \centering
    \includegraphics[width=\columnwidth]{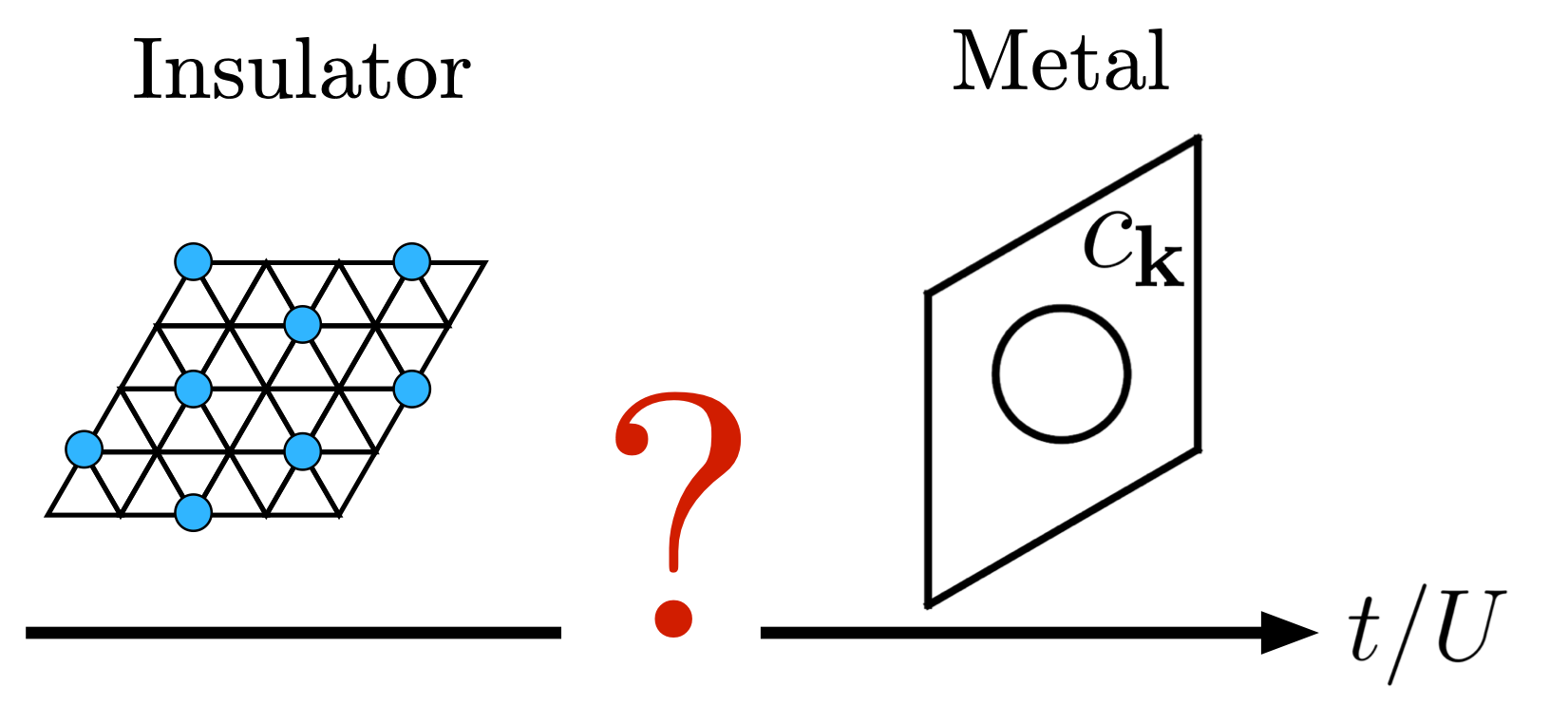}
    \caption{A `Wigner-Mott' (WM) insulator with charge-density wave order evolves into a metal as a function of increasing bandwidth at fixed fractional filling. This paper analyzes critically the possibility of a direct continuous transition between a (different) WM insulator and a metal with an electronic Fermi surface.}
    \label{fig:phase_trans0}
\end{figure}

\begin{figure*}
    \centering
    \includegraphics[width=\textwidth]{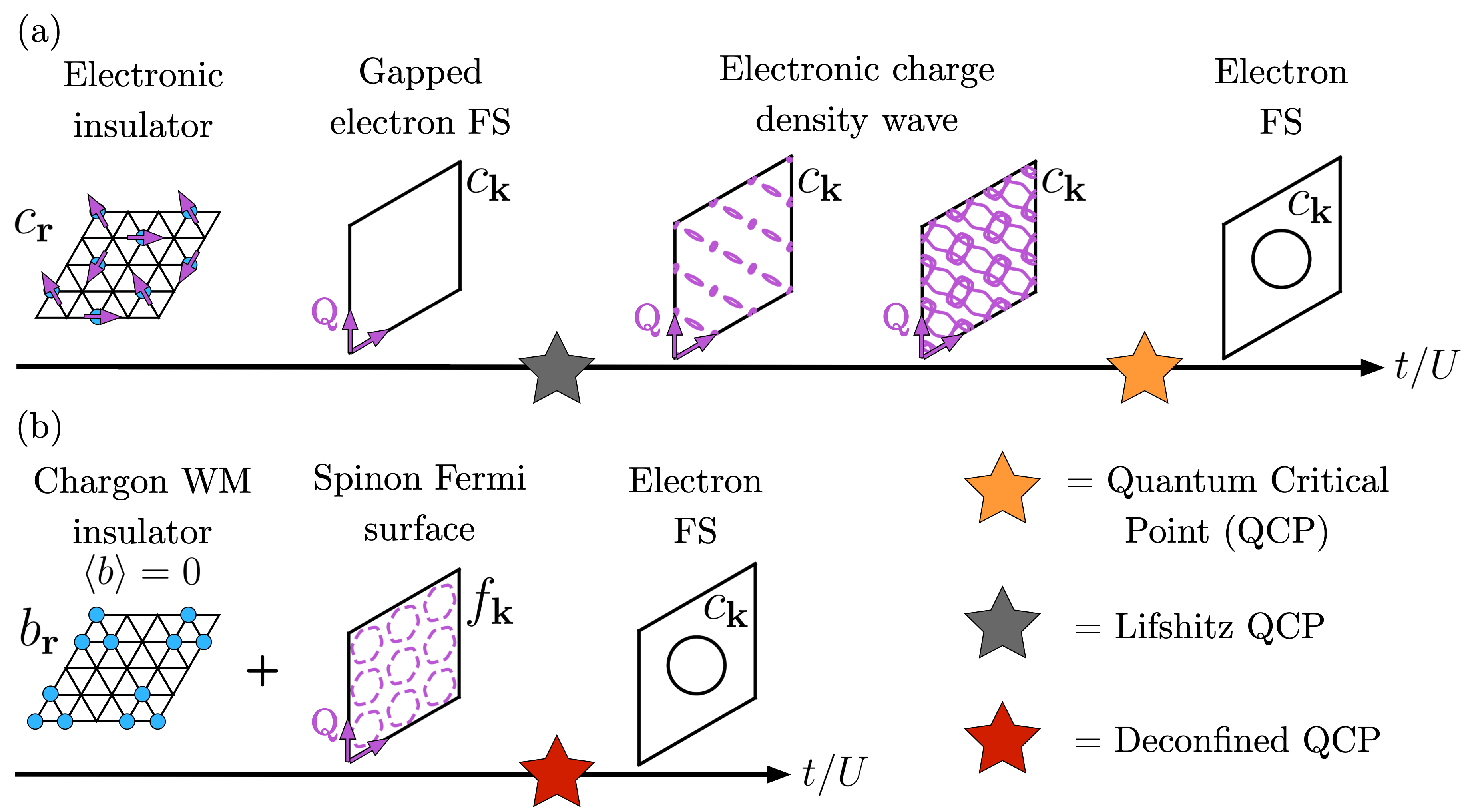}
    \caption{Scenarios for continuous MIT at {\it fixed} $\nu_c = 1/q$ (for $q>2$): (a) A conventional transition where the electronic Fermi surface is reconstructed by spin-density wave order and shrinks continuously with increasing strength of this order via a Lifshitz transition. This type of transition is discussed in detail in Sec. \ref{sec:prelimHF}. (b) The main subject of the present paper concerns a direct transition between the WM insulator and a metal without any spontaneously broken symmetries. The chargon theory is then described by a `deconfined' critical point between a superfluid and charge ordered Mott insulator. Within our theory the charge order will not be of the $\sqrt{3}\times \sqrt{3}$ type; this is discussed in the main text. We show instead the triangular superlattice which is a possible mean field charge order in our theory.}
    \label{fig:phase_transition}
\end{figure*}

In this paper, we will be concerned with the important theoretical question regarding the nature of the evolution from a symmetry-preserving Fermi liquid to a  WM insulator, which has co-existing crystalline orders, at fixed $\nu_c ~ (<1/2)$. This is illustrated in Fig.~\ref{fig:phase_trans0}. The metal has an electronic  Fermi surface with an area (in units of the Brillouin zone area) fixed by Luttinger's theorem by the filling $\nu_c$. The WM insulator on the other hand has no electronic Fermi surface but broken space-group symmetries. Given the striking differences between these two phases an obvious expectation for this evolution is that it occurs through a direct first-order transition. Alternately the evolution can also happen in a more elaborate manner through intermediate phases which are charge ordered but remain metallic. Both of these scenarios have been considered before in numerous other works, such as \cite{imada_metal-insulator_1998, jamei_universal_2005, camjayi_coulomb_2008, amaricci_extended_2010}. The latter is realized in a simple weak-coupling treatment of models of correlated electrons. In the intermediate phase the broken translation symmetry leads to an even number of electrons in the enlarged unit cell. The electronic Fermi surface can then be shrunk continuously to zero across the metal-insulator transition. A transition of this type is illustrated in Fig.~\ref{fig:phase_transition}(a) and discussed in Sec.~\ref{sec:prelimHF}. Previous works using a variety of mean-field-like approaches have studied these types of transitions in twisted bilayer transition metal dichalcogenides \cite{pan_interaction-driven_2021, zang_hartree-fock_2021, pan_interaction_2021}.

But is a direct continuous  transition between the symmetry-preserving Fermi liquid and a Wigner-Mott insulator at all allowed even in principle? Contrary to the conventional scenarios described in the previous paragraph, such a transition must necessarily involve both the continuous disappearance of the full Fermi surface (without shrinking), and the concomitant development of charge order which breaks space-group symmetry.  Importantly,  the onset of the broken symmetry alone cannot account for the disappearance of the entire FS across the transition into the insulator. Thus, in spite of the existence of a possible order parameter associated with the spontaneously broken translational symmetry, the continuous metal-insulator transition at fixed $\nu_c$ lies fundamentally beyond a mean-field order-parameter-based paradigm in the absence of any fine-tuning (e.g., in the form of FS `nesting'). As a result, the low-energy field theory will  not be governed by the fluctuations of these order parameter fields and necessitates a more complex description. 

Remarkably we will show in this paper that continuous Wigner-Mott transitions are indeed possible. We describe a low-energy effective field theory for such transitions  and determine many of its universal critical properties. The WM insulator we find will be fractionalized with a Fermi surface of electrically neutral spinons.

Our results generalize the theory of continuous Mott transitions \cite{senthil_theory_2008} to fillings $\nu_c < 1/2$ where WM insulators can arise. To that end we will work with a parton description where the electron operator at site $r$ and spin $\alpha$ is fractionalized as a product: $c_{r \alpha} = b_r f_{r\alpha}$ where $b_r$ is a spinless boson (the chargon) that carries the electric charge of the electron, and $f_{r\alpha}$ is an electrically neutral  spin-$1/2$ fermion (the spinon). In this  representation  the conventional Fermi liquid is described \cite{lee_doping_2006} as a superfluid of chargons in the presence of a Fermi surface of spinons. The WM insulator, on the other hand, is described  as a bosonic WM chargon insulator in the presence of the spinon Fermi surface. The transition between these two phases is thus viewed as a superfluid-WM insulator transition of the bosonic chargons.   A direct continuous transition between these two phases of bosons is forbidden by standard Landau theory as they break distinct symmetries. However Landau-forbidden continuous phase transitions are known to occur, and are described by the theory of deconfined quantum criticality \cite{senthil2004deconfined,senthil2004quantum}. We will thus first study the possibility of such a Landau-forbidden superfluid-WM transition of bosons. For the triangular lattice (which will be our main concern here) a theory for such a transition was formulated in Ref. \onlinecite{burkov_superfluid-insulator_2005} using a dual description in terms of vortex fields. We will introduce a large-$N$ generalization of this model that allows us to calculate its properties. We will then include the coupling to spinons (and associated emergent gauge fields) to study the electronic WM transition. A cartoon of this transition is illustrated in Fig.~\ref{fig:phase_transition}(b).

A recent work \cite{xu_metal-insulator_2021} has also discussed the possibility of a continuous MIT, where the insulator breaks translational symmetry, focusing mostly on the case of half filling on the triangular lattice. Importantly, the authors do not address the role of the possibly relevant couplings between gauge-invariant composite operators in the chargon and spinon sectors, respectively. In the absence of a ``dynamical decoupling" between these matter sectors (Ref.~\cite{senthil_theory_2008}; see below), the resulting low-energy theory can become strongly coupled and the transition need not be described solely in terms of condensation of the chargon fields. We analyze this aspect in the present paper carefully, finding a solvable example where the chargon and spinon sectors decouple dynamically at the critical point.

The remainder of this article is organized as follows: In Sec.~\ref{sec:prelimHF}, we illustrate through several examples the inherent challenges associated with constructing a simple order-parameter-based Hartree-Fock type theory for describing a continuous transition between a metal and a WM insulator at fixed electron filling (in the absence of fine-tuning). In Sec.~\ref{sec:prelimreview}, we review the key features associated with electron ``fractionalization" and a parton-based framework that allow us to describe such continuous MIT at fixed electron filling in the simpler setting of $\nu_c=1/2$. In Sec.~\ref{sec:WM_cont}, we use the parton formulation and the charge-vortex duality to develop a theory for the metal to WM insulator transition at fixed $\nu_c=1/q$. Section \ref{sec:expt} is devoted to discussing the salient experimental signatures near the MIT and we end with an outlook toward a number of pressing questions in Sec.~\ref{sec:outlook}. A number of technical details are summarized in the appendices.

\section{Preliminaries}

\subsection{Limitations of a weak-coupling analysis}
\label{sec:prelimHF}

\begin{figure}
    \centering
    \includegraphics[width=\columnwidth]{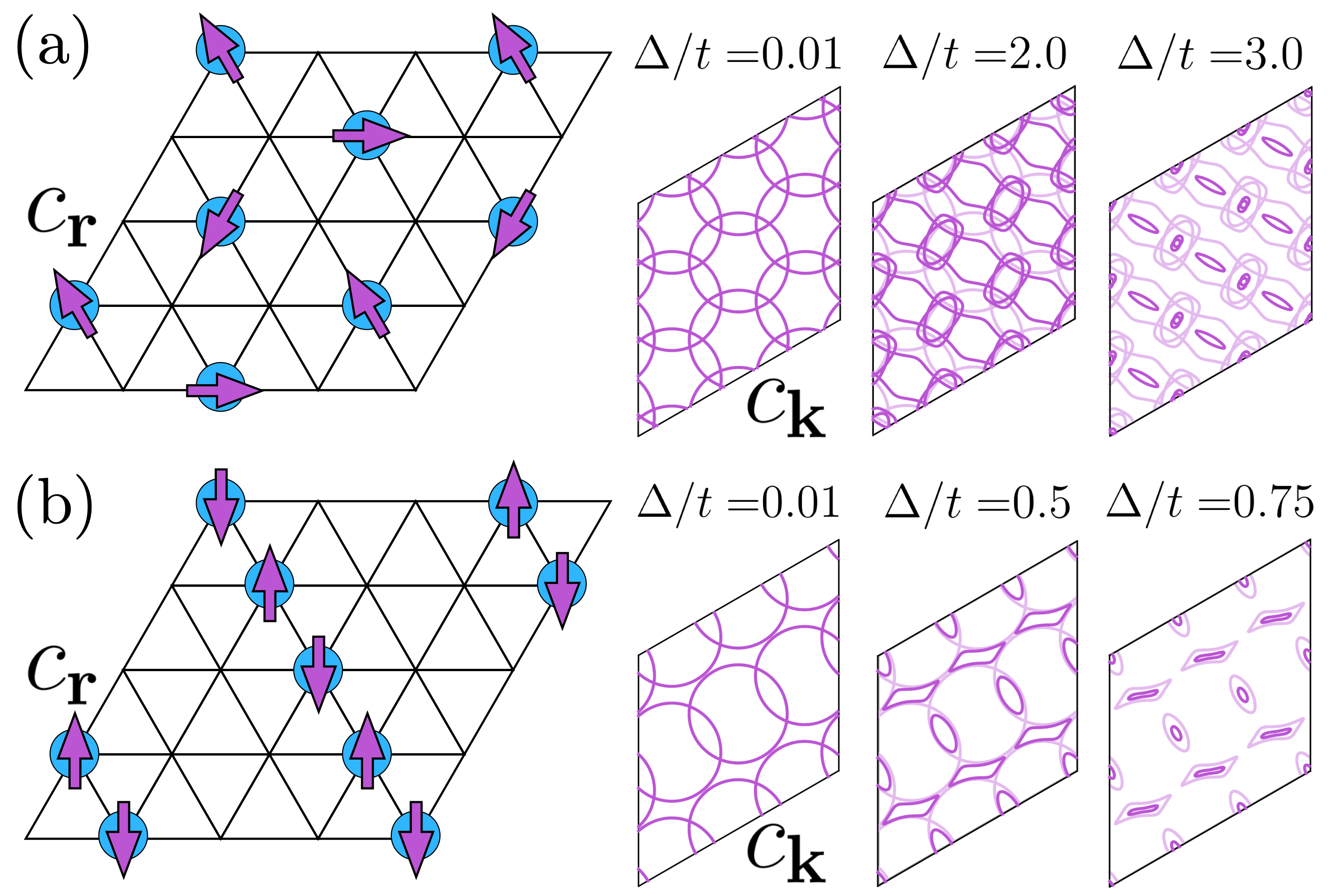}
    \caption{Evolution of the electronic Fermi surface with increasing magnitude of various mean-field order parameters. (a) Left: The proposed WM insulator is shown in real space, where the blue circles indicate electronic density and the arrows indicate the spin configuration. Right: The evolution of the electronic Fermi surface in the original Brillouin zone as a function of increasing $\Delta/t$, where $t$ is proportional to the bandwidth. The Fermi surface disappears around $\Delta/t \gtrsim 4$. The faint traces for the Fermi surface correspond to smaller values of $\Delta/t$ and are overlayed in each panel for tracking the shrinking evolution. (b) Left: The proposed WM insulator in real space with the same notation as in (a). Right: The evolution of the electronic Fermi surface with increasing $\Delta/t$. The Fermi surface disappears for $\Delta/t\gtrsim 1$.}
    \label{fig:HF}
\end{figure}

In this section, we will discuss two examples of the onset of simple order parameters in a metal, inspired by the possible density-wave states in the WM insulator at a fixed $\nu_c=1/6$ on the triangular lattice. As expected, in both cases the generic electronic Fermi surface will not disappear immediately upon the spontaneous breaking of translational symmetry. Instead, the phase across the critical point describes a metal with reconstructed electronic Fermi surfaces. These reconstructed Fermi surfaces shrink and can eventually be gapped out once the magnitude of the order parameter becomes large. More generally, a generic electronic Fermi surface cannot be gapped entirely due to an order parameter that carries a finite center-of-mass momentum across a continuous transition, regardless of its precise microscopic nature.

Let us consider a $\sqrt{3} \times \sqrt{3}$ electronic charge-ordered state pictured in Fig.~\ref{fig:phase_trans0}; note that the electrons have a remaining spin degree of freedom, which needs to be accounted for in order to describe an insulating state. Starting from the long-ranged Coulomb interactions in the microscopic model, the magnetic ordering in the insulating state will be determined by the competition between various ferro- and antiferromagnetic exchange interactions \cite{pan_interaction-driven_2021}. A possible state with the spins forming a $120^{\circ}$ N\'{e}el order on the effective triangular lattice ($\sqrt{3}\times \sqrt{3}$) Mott insulator is shown in Fig.~\ref{fig:HF}(a). These orderings can be readily implemented at a mean-field level (see Appendix \ref{app:HF_120}). As the strength of the magnetic (spin density wave) order, $\Delta$, is increased continuously from zero, there is an onset of a charge gap proportional to $\Delta$ along certain portions of the electronic Fermi surface. At fixed $\nu_c=1/6$, as $\Delta$ increases to the order of the single-electron bandwidth, the system eventually becomes a robust band insulator (Figure~\ref{fig:HF} (a)). As noted earlier, the electronic Fermi surface does not disappear instantly with the onset of the order, but shrinks to zero via a Lifshitz-type transition deep inside the magnetically ordered metallic state.

It is possible to host other forms of translational symmetry breaking at the same filling; consider, e.g., the case of an electronic `stripe' order (Figure~\ref{fig:HF}(b)). The charge-density with period three arranges itself into unidirectional stripes and we must further address the fate of the spin to account for an insulating phase. A possible state is shown in Figure~\ref{fig:HF}(b), where the spins order antiferromagnetically along the unidirectional stripes. When this density wave is implemented, there are six sites per unit cell, allowing for a band insulator at filling $\nu_c=1/6$. As discussed in Appendix \ref{app:HF_stripe}, once the strength of the density wave becomes of order the single-electron bandwidth, the system will again become a robust band insulator at filling $\nu_c=1/6$. However, as in the previous example, the Fermi surface will not disappear instantly with the onset of the combined charge and spin density waves, but will rather undergo a Lifshitz transition out of a metallic density-wave state.

\subsection{Review of continuous MIT at $\nu_c=1/2$}
\label{sec:prelimreview}

To describe a continuous MIT at fixed fractional (commensurate) fillings, we will begin by writing the electron operator in terms of fractionalized degrees of freedom (i.e., partons) coupled to emergent, dynamical gauge fields. The resulting field theory in terms of these new degrees of freedom is strongly coupled and requires careful analysis. We will briefly review the key elements of the theory which describes a transition from a metal to a paramagnetic Mott insulator with a spinon Fermi surface at $\nu_c=1/2$, discussed in more detail in \cite{senthil_theory_2008}, before generalizing it to other $\nu_c=1/q~(q>2)$ in Sec.~\ref{sec:WM_cont} below. At the outset, we note that a field theory for our transition of interest has not been studied before, even in the setting of the density (or $r_s$) tuned MITs that have been analyzed extensively in the context of the two-dimensional electron gas.  

We write the electron operator as $c_{\r\alpha} = b_{\r} f_{\r\alpha}$, where $b_{\r}$ is a spinless, charge-$e$ boson (`chargon') and $f_{\r\alpha}$ is a spin$-1/2$ electrically neutral fermion (`spinon'). In the simplest theory, both of these fields are coupled minimally to an emergent dynamical $U(1)$ gauge field, $a_{\r}\equiv(a_0,\a)$. The general form of the imaginary-time action for the matter fields at any filling can then be written as
\begin{subequations}
\beq
\S &=& \S_{[f,a]} + \S_{a} + \S_{[b,a]} + \S_{[b,f]}, \label{FT}\\
\S_{[f,a]} &=& \int_{\tau,\k}f_{\k\alpha}^\dagger (\partial_\tau - i a_0 - \ve^f_{\k-\a} + \mu_f) f_{\k\alpha},\\
\S_a &=& \frac{1}{2e^2_a} f^2_{\mu\nu}, \\
\S_{[b,f]} &=& \int d\tau~d^2\r~ \O_b \O_f, \label{Sbf}
\eeq
\end{subequations}
where $(\ve^f_\k - \mu_f)$ represents the spinon dispersion (including the chemical potential, $\mu_f$) and $f_{\mu\nu} = \partial_\mu a_\nu - \partial_\nu a_\mu$ is the field strength, with $e$ the gauge coupling. The terms in $\S_{[b,f]}$ include all gauge-invariant operators, $\O_b$ and $\O_f$, respectively; a specific example includes the chargon energy-density ($\O_b = |b|^2$) coupled to a fermion bilinear ($\O_f = f^\dagger f$). For the case of the WM transitions considered later, special care will be needed to address the fate of couplings to a number of physically relevant $\O_b$.

For the bandwidth-tuned MIT at $\nu_c=1/2$, $\S_{[b,a]}$ is described by the theory for a relativistic boson, $b$, coupled minimally to the gauge field, $a$:
\beq
\S_{[b,a]} = \int_{\tau,\r}~ \bigg[|(\partial_\mu - ia_\mu) b|^2 + s|b|^2 + \lambda |b|^4 + ... \bigg].
\eeq
At the mean-field level, the MIT is tuned by driving a superfluid ($\la b\ra\neq0$) to Mott insulator ($\la b\ra=0$) transition for the boson. The transverse components of the gauge field $\a$ receive singular frequency and momentum-dependent corrections from the gapless matter fields (at the critical point), leading to a $z=2$ dynamics. However, the feedback of $\a$ on the boson dynamics remains non singular and can be ignored; as a result, $\S_{[b,a]}$ is effectively described by a 3D-XY transition for $b$ in the presence of a spinon Fermi surface coupled to $\a$. The fate of the MIT and the low-energy field theory is then dictated solely by $\S_{[b,f]}$. For the energy-energy coupling considered earlier, as long as the correlation length exponent $\nu>2/3$, the coupling is irrelevant. This is indeed the case for the 3D-XY transition. Therefore at $\nu_c=1/2$, the $b$ and $f+\a$ sectors of the theory decouple {\it dynamically} from each other. This is crucial for describing the fate of the ultimate low-energy properties of the resulting theory, ${\cal{S}}$, and the complex temperature-dependent crossovers in the vicinity of the MIT. The critical point also hosts a sharp {\it electronic} Fermi surface without any low-energy quasiparticles --- a `critical Fermi surface'\cite{senthil2008critical}.

\section{Continuous Wigner-Mott transition at $\nu_c=1/q$}
\label{sec:WM_cont}

Let us now generalize the action, ${\cal{S}}$, for describing continuous bandwidth-tuned transitions at other fixed commensurate fillings. Specifically, the important modifications will arise in the form of $\S_{b}$ and $\S_{[b,f]}$, respectively.

\subsection{Criteria for dynamical decoupling}
\label{sec:criteria}

As noted earlier, we will be interested in the second scenario in Fig.~\ref{fig:phase_transition}(b), where the chargons undergo a superfluid to WM insulator transition with broken translational symmetry. However, before specifying the action $\S_b$ that can describe such a transition, it is important to note a complication that can arise due to the onset of broken translational and/or rotational symmetries in the WM insulator. The ultimate fate of the continuous MIT at a fixed fractional $\nu_c$ is governed by $\S_{[b,f]}$. In addition to the possible terms that were already considered above (e.g., energy-density couplings), the presence of various crystalline orders can complicate the nature of the transition. 

As was already evident in the discussion of Fermi surface reconstruction across a density-wave ordering in Sec. \ref{sec:prelimHF}, the interplay of the order parameters that describe the onset of various crystalline point-group symmetry breaking and the excitations near the gapless Fermi surface can lead to a non-trivial dynamics at low energies. The same considerations will also apply near the MIT, where the electron Fermi surface disappears and evolves into the spinon Fermi surface; see Fig.~\ref{fig:phase_transition}(b). Therefore, we need to include a coupling between the particle-hole fluctuations near the spinon Fermi surface to $\O_b\equiv \rho_\Q,~\N$, where $\rho_\Q$ is a charge-density wave order at wave vector $\Q$ and $\N$ is a nematic order (associated with a spontaneously broken lattice rotational symmetry) in $\S_{[b,f]}$.

Integrating out the low-energy fermionic excitations near the spinon Fermi surface generates terms in the effective action in imaginary frequency of the form,
\beq
\S_{\tn{eff}}[\rho_\Q,\N] = \int_{\omega,\q}\bigg[|\omega| |\rho_\Q|^2 + \frac{|\omega|}{q}|\N|^2 \bigg], \label{Scdw}
\eeq
as a result of familiar Landau damping. In order for $\S_{\tn{eff}}$ to be an irrelevant perturbation at the critical point between the metal and WM insulator, we require that the anomalous dimension for the charge-order and nematic fields satisfy $\eta_\rho>1$ and $\eta_\N>2$, respectively. If satisfied, the different matter-field sectors (i.e., the chargons and spinons) will once again decouple dynamically, provided the energy-energy couplings are also irrelevant, just as in the $\nu_c=1/2$ MIT. Note that the above constraints on $\eta_\rho$ and $\eta_\N$ do not depend on the precise microscopic relationship between the order parameters and the chargon fields, which we will specify explicitly in Sec.~\ref{sec:DD}. However, the actual values of $\eta_\rho,~\eta_\N$ will depend on $\S_b$ that is appropriate for a given MIT. Importantly, $\S_b$ must describe a continuous superfluid to Wigner crystal transition of the chargons, which is Landau forbidden. The theory for the chargons must necessarily be a deconfined quantum phase transition. In the next few sections, we introduce a theory that realizes such a deconfined phase transition for the chargons and analyze its fate in the presence of coupling to the gapless spinons.

\subsection{Dual vortex theory}
\label{sec:dual}

Instead of working with the chargon ($b$) fields directly, it will be fruitful to consider the dual vortex theory \cite{lannert_quantum_2001}, where the vortex field is coupled to a non compact gauge field, $A$, dual to the Goldstone mode of the superfluid. The vortex condensate is the bosonic Mott insulator and the vortex insulator is the bosonic superfluid. Importantly, when a vortex winds around a site containing a boson it will pick up a flux from $A$. When the boson density is $\nu_b = 2\nu_c$, the flux upon winding around a lattice site is $2\pi \nu_b$. The resulting vortex multiplet theory then necessarily transforms under the projective symmetry group (PSG) of the underlying lattice.

\begin{figure}
    \centering
    \includegraphics[width=\columnwidth]{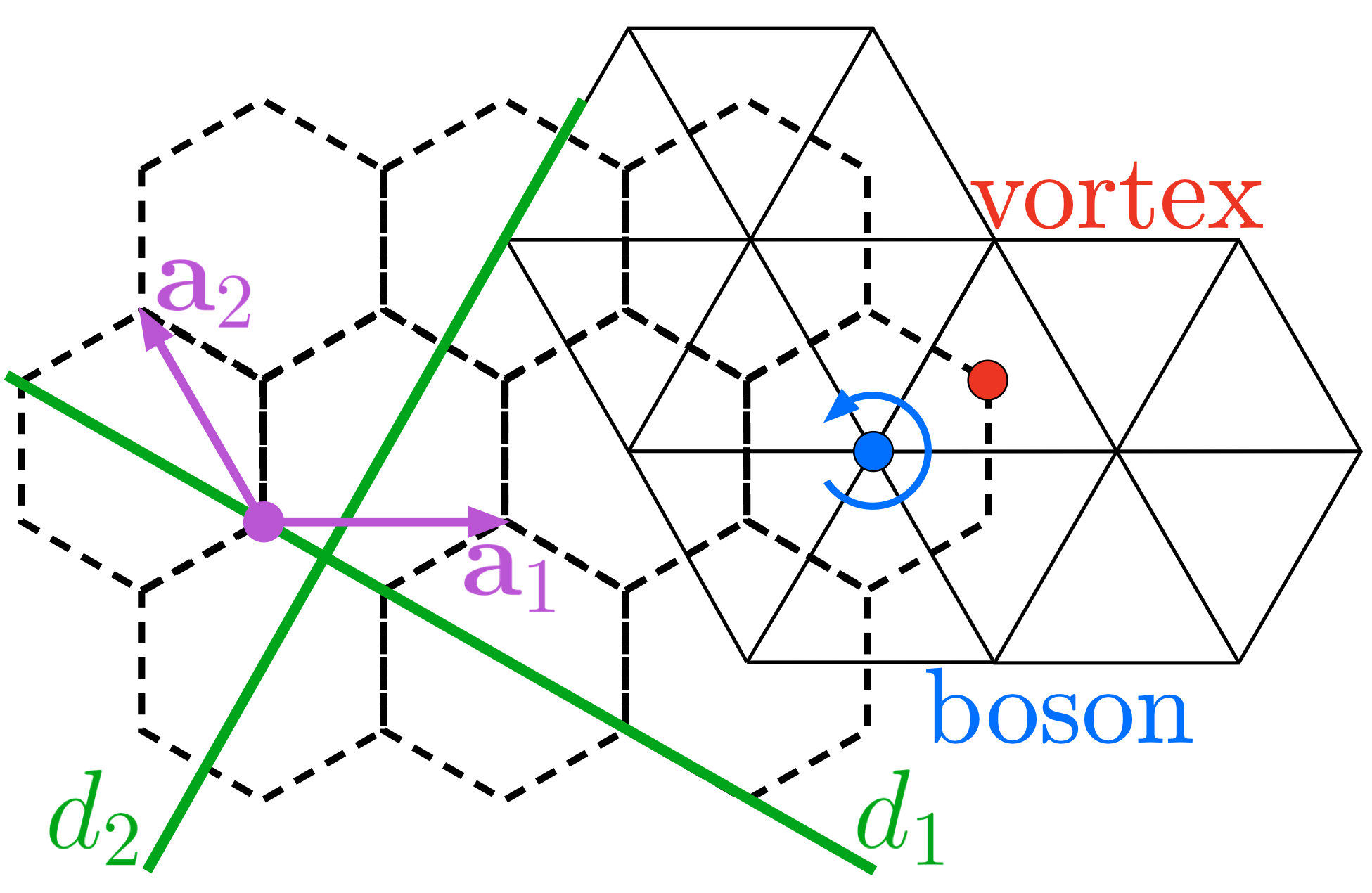}
    \caption{The triangular lattice with bosons in blue and vortices in red. Vortices live in the center of the triangular cells, or alternatively on the sites of the dotted dual hexagonal lattice. Each boson will induce a phase when a vortex path encloses it. Also shown are the generators of the triangular symmetry group: $T_{1,2}$ are translations by $\mathbf{a}_{1,2}$; $I_{d_1},I_{d_2}$ are reflections about $d_1,d_2$; and $R_{2\pi/3}$ is a rotation by $2\pi/3$ about the site labeled with a magenta dot.}
    \label{fig:d_lat}
\end{figure}

In order to be concrete, we will focus on the case of $\nu_c=1/6$ and describe a MIT from a metal to WM insulator with charge order. On the triangular lattice, which is of special interest in light of the moir\'e TMD experiments, $\S_{b}$ can be expressed in terms of the vortex multiplet, $\phi_l$ ($l=0,1,2$), as
\beq
\S_{[b,a]} &=& \int_{\tau,\r} \bigg[\sum_{l} |(\partial_\mu - iA_\mu) \phi_l|^2 + s|\phi_l|^2 \nonumber\\
&+&  \lambda\bigg[\sum_l|\phi_l|^2\bigg]^2 + g\sum_l|\phi_l|^4 + \cdots\nonumber\\
&+& \frac{1}{2e^2_A}\left(\epsilon^{\mu\nu\lambda}\partial_\nu A_\lambda\right)^2 + \frac{1}{2\pi} \epsilon^{\mu \nu\lambda}a_\mu\partial_\nu A_\lambda\bigg]. \label{eqn:sbWM}
\eeq
This action is discussed in much more detail in \cite{burkov_superfluid-insulator_2005}. The vortex multiplet, $\phi_l$, represents the ``permutative representation" of the PSG transformations, as listed in Table \ref{tab:PSG} (and see Fig.~\ref{fig:d_lat}).\footnote{These variables are denoted $\xi_l$ in Ref.~ \cite{burkov_superfluid-insulator_2005}.} These vortex flavors can be thought of as tied to fractional chargons. Indeed if $n$ of the vortex flavors are condensed then a ``vortex" of any of the condensed $\phi_l$, i.e. a state where the phase of any $\phi_l$ winds by $\pm 2\pi$ at infinity, will carry an attached flux of $2\pi/n$ and will thus correspond to a localized boson number of $\pm 1/n$. The emergent gauge field thus couples to the theory via a mutual Chern-Simons term with $A_\mu$, as $\epsilon^{\mu \nu \lambda}\partial_\nu A_\lambda$ represents the physical chargon current.

We have included up to fourth-order terms in the expansion above, consistent with the PSG transformations. In particular, the equations are consistent with the full permutation symmetry of the vortex multiplets implied by the triangle PSG and with time-reversal symmetry. Additionally they possess particle-hole symmetry for each vortex multiplet. It is worth noting that $\S_b$ in Eq.~(\ref{eqn:sbWM}) is also invariant under an emergent internal symmetry $U(1)^3 /U(1)$, which is broken by a term at the sixth order that is consistent with the PSG transformations: 
\beq
\L_w = w\bigg[\left(\phi_0^* \phi_1\right)^3 + \left(\phi_1^*\phi_2\right)^3 + \left(\phi_2^*\phi_0\right)^3 + \mathrm{h.c.}\bigg].
\label{eqn:w_term}
\eeq
At the mean-field level, $\S_b$ in Eqn.~\ref{eqn:sbWM} can be analyzed readily. If $g<0,~ \lambda > |g|$ the ground state for $s<0$ corresponds to having only one of the vortex flavors condensed, leading to an insulator with a single chargon on every third site 
\cite{burkov_superfluid-insulator_2005}; see Fig.~\ref{fig:phase_trans0}. This is the Wigner crystalline phase that one would expect from purely classical considerations due to the further neighbor Coulomb repulsion on the triangular lattice, and is quite likely the phase observed in certain moir\'e TMD bilayers at strong interactions \cite{li_imaging_2021}. On the other hand, for $g>0$ it will be energetically favorable to condense all of the vortex flavors, leading to other charge-ordered insulators \cite{burkov_superfluid-insulator_2005} (see Fig.~\ref{fig:RG_flow}(a)). As discussed above, in these states the chargon will generically split into three.

The role of fluctuations beyond the mean-field level and the associated low-energy properties of $\S_b$ are not presently known. To make controlled analytical progress and examine the key theoretical issues that determine the nature of the MIT, including the fate of dynamical decoupling, we will construct a `solvable' large$-N$ limit in the next subsection, where the low-energy properties can be worked out reliably.

\begin{table}
\centering
 \begin{tabular}{||c||c|c|c||} 
 \hline
   & $\phi_0$& $\phi_1$& $\phi_2$\\
 \hline\hline
 $T_1$& $\phi_1$& $\phi_2$& $\phi_0$\\
 $T_2$& $\omega \phi_1$& $\omega^2 \phi_2$& $\phi_0$\\
 $R_{2\pi/3}$& $\omega^{1/4}\phi_2$& $-i\phi_0$& $\omega^{1/4}\phi_1$\\
 $I_{d_1}$& $\omega^{-1/4}\phi_2^*$& $-\omega^{1/4}\phi_1^*$& $\omega^{-1/4}\phi_0^*$\\
 $I_{d_2}$& $\omega^{-1/4}\phi_0^*$& $-\omega^{1/4}\phi_1^*$& $\omega^{-1/4}\phi_2^*$\\
 \hline
 \end{tabular}
 \caption{Triangle PSG consistent with the gauge choice for $A$ made in \cite{burkov_superfluid-insulator_2005}. Here $\omega = e^{2\pi i/3}$ and the threefold translational symmetry is evident. These generators are pictured in Fig.~\ref{fig:d_lat}.}
 \label{tab:PSG}
\end{table}

\subsection{Large$-N$ extension of the dual vortex theory} 
\label{sec:largeN}

Let us promote each of the vortex flavors to complex vectors with $N$ components ($\in \bb{C}^N$): $\{\phi_0,\phi_1,\phi_2\} \rightarrow \{{\vec\phi}_0,{\vec\phi}_1,{\vec\phi}_2\}$, and promote the action in Eq.~(\ref{eqn:sbWM}) to 
\beq
\S_b \rightarrow&&\int_{\tau,\vec{r}} \bigg(\left|\left(\partial_\mu - \frac{i}{\sqrt{3N}}A_\mu\right)\bs{\phi}\right|^2 + s|\bs{\phi}|^2 + \lambda \left(|\bs{\phi}|^2\right)^2 \nonumber\\
&+& g\sum_{l=0}^{2} \left(|\bs{\phi}_l|^2\right)^2 +  \frac{1}{2e^2_A}\left(\epsilon^{\mu\nu\lambda}\partial_\nu A_\lambda\right)^2 \bigg) \label{eqn:large_N},
\eeq
where $\bs{\phi} = (\bs{\phi}_0, \bs{\phi}_1, \bs{\phi}_2)\in \bb{C}^{3N}$.{\footnote{This theory has an internal $SU(3N)$ symmetry which is broken to a $U(N)^3 \rtimes S_3/U(1)$ symmetry by the $g$ term. Note that the PSG transformations act trivially on the additional degrees of freedom. The $w$ term in Eq.~(\ref{eqn:w_term}) will further break this to a $U(N) \rtimes S_3/U(1) \simeq SU(N) \rtimes S_3$ symmetry. The large-$N$ extension displays an extra global $SU(N)/\bb{Z}_N$ symmetry; our ultimate interest is in the physical $N=1$ limit of this construction. The $SU(N) \rtimes S_3$ symmetry group preserved by the $w$ term is the smallest internal symmetry group that is consistent with the both the permutation required by the triangle PSG in Table \ref{tab:PSG} and the requirements for global $SU(N)$ symmetry.}} Note that we have rescaled the noncompact $U(1)$ gauge field $A_\mu$ so that it will properly drop out in the $N\rightarrow \infty$ limit; see \cite{kaul_quantum_2008}.

\begin{figure}
    \centering
    \includegraphics[width=\columnwidth]{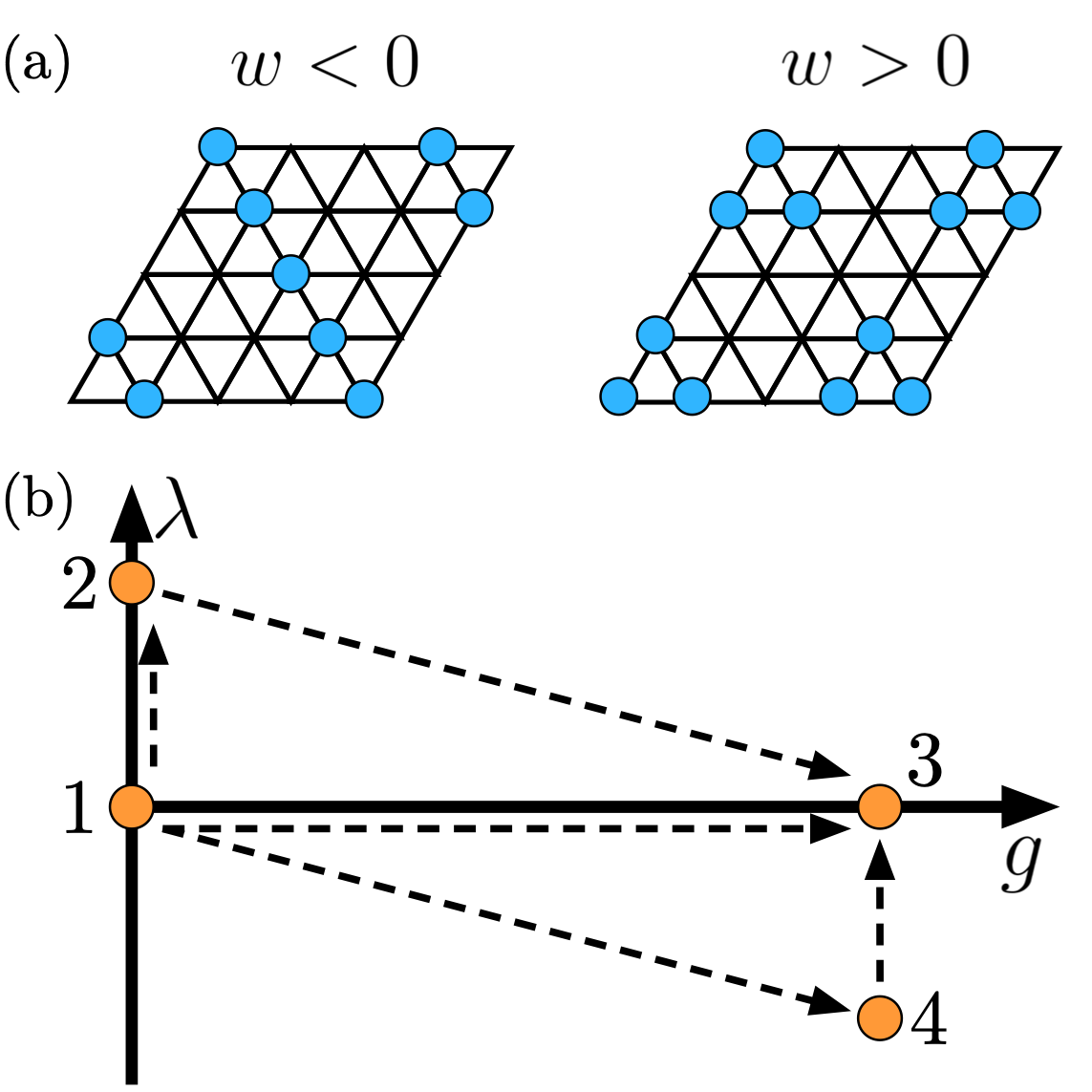}
    \caption{(a) Possible mean-field states for $g>0$. If $w<0$ the system will prefer stripe order in one of the three possible stripe directions, while if $w>0$ the system will prefer clusters of chargons that form a triangular superlattice. Note that Eq.~(\ref{eqn:sbWM}) and Eq.~(\ref{eqn:large_N}) have an additional singlet term $|\bs{\phi}|^8$ with a positive coefficient that ensures stability even if $w<0$. (b) The RG flow in the $N=\infty$ limit of Eq.~(\ref{eqn:large_N}). In this limit the stable fixed point corresponds to $\lambda =0$, $g>0$, or three decoupled $O(2N)$ models. The charge configuration in the mean field will thus be one of the two possibilities in (a).}
    \label{fig:RG_flow}
\end{figure}

In this limit the theory then has an emergent internal $O(6N)$ symmetry which is broken to $O(2N)^3 \rtimes S_3$ due to the term proportional to $g$. Based on a $(4-\epsilon)-$expansion   \cite{osborn_seeking_2018}, the theory has the following four fixed points (Fig.~\ref{fig:RG_flow}(b)): 
\begin{enumerate}
\item[(1)] the Gaussian fixed point with $g = \lambda = 0$,
\item[(2)] the fully $O(6N)$ symmetric theory with $g=0$, $\lambda >0$, 
\item[(3)] three decoupled $O(2N)$ models with $\lambda=0$, $g>0$, and 
\item[(4)] the $O(2N)^3 \rtimes S_3$ fixed point with $\lambda,g\neq 0$ \cite{henriksson_perturbative_2021}. 
\end{enumerate}
Furthermore the stable fixed point will be (3), i.e.m when $\lambda=0$ and $g>0$ \cite{osborn_seeking_2018}, implying that the $N\rightarrow\infty$ theory will condense all vortex flavors at the mean-field level on the insulating side. Note that the resulting charge-density wave order will not be given by the usual $\sqrt{3}\times\sqrt{3}$ pattern, as in Fig.~\ref{fig:phase_trans0}. Instead, the mean field charge-density patterns will be as shown in Fig.~\ref{fig:RG_flow}(a), depending on the sign of the $w$ term in Eq.~(\ref{eqn:w_term}). Within our present large$-N$ framework, we will investigate next whether a continuous MIT is possible between a metal without any broken symmetries and either of the two ordered WM insulators shown in Figure~\ref{fig:RG_flow}(a). This primarily requires addressing the fate of dynamical decoupling, as determined by the terms contained in $\S_{[b,f]}$ (Eq.~\ref{Sbf}) and $\S_{\tn{eff}}[\rho_\Q,\N]$ [Eq.~(\ref{Scdw})]. Note that at the fixed point (3) all terms that are sixth order or higher, e.g., the $w$ term, will be marginally irrelevant in $(2+1)-$ dimensions.

\subsection{Dynamical decoupling and fate of continuous MIT}
\label{sec:DD}

In the $N\rightarrow\infty$ limit, for the fixed point denoted (3) in the previous subsection and Figure~\ref{fig:RG_flow}(b), the correlation length exponent
\beq
\nu = 1 - \mathcal{O}\left(\frac{1}{N}\right),
\eeq
and the singlet operator thus has scaling dimension $\Delta_S = d - 1/\nu = 2 + \mathcal{O}(1/N)$. At the decoupled $O(2N)$ critical point the singlet operator is any linear combination of the $|\bs{\phi}_l|^2$. All other operators have the scaling dimensions they exhibit at the Gaussian fixed point, i.e., the na\"{i}ve scaling dimension \cite{moshe_quantum_2003}. This allows us to immediately dispatch with the energy-energy couplings expressed in $\S_{[b,f]}$. Based on the arguments we reviewed in Sec.~\ref{sec:prelimreview}, such couplings will be irrelevant at the critical point \cite{senthil_theory_2008} since $\nu = 1>2/3$ in the $N=\infty$ limit. The only thing remaining then is to investigate whether the effective action that is generated in Eq.~(\ref{Scdw}) is irrelevant at the same critical point.

\begin{figure}
    \centering
    \includegraphics[width=\columnwidth]{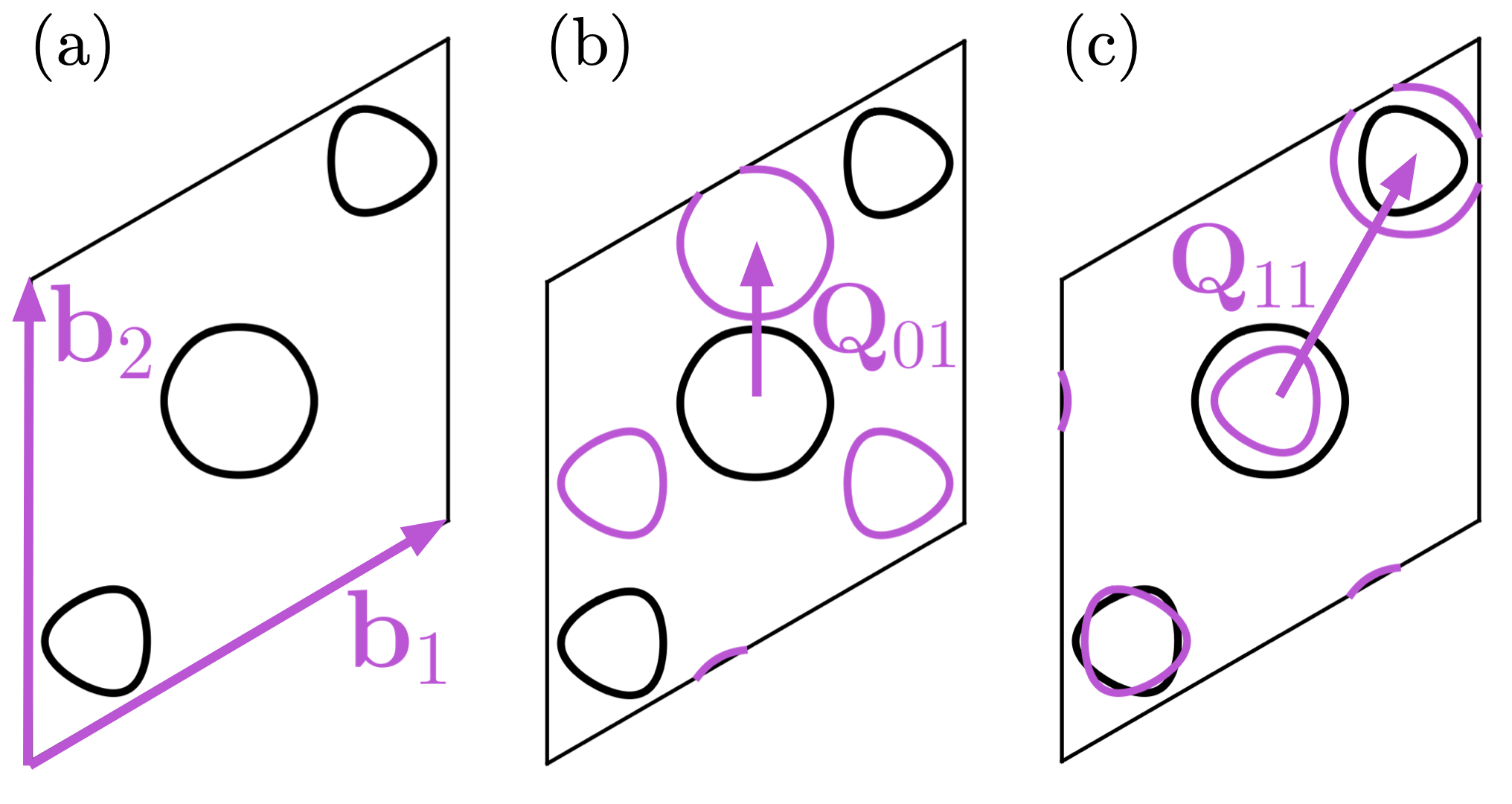}
    \caption{The Fermi surface at $\nu_c = 1/6$ produced by a third-nearest-neighbor hopping model on the triangular lattice with $t_2/t_1 = 2, t_3/t_1=-1.25$. (a) The Fermi surface is shown in black. The reciprocal lattice vectors $\vec{b}_1, \vec{b}_2$ are also pictured in purple. (b) The Fermi surface shifted by the wave vector $\vec{Q}_{01} = \vec{b}_2/3$ is shown in purple; it avoids hot spots with the original Fermi surface. Though not shown this Fermi surface also avoids hot spots for all $\vec{Q}_{mn}$ with $m\neq n$. (c) The Fermi surface shifted by $\vec{Q}_{11} = (\vec{b}_1+\vec{b}_2)/3$ is shown in purple; parts of it are nearly nested with the original Fermi surface.}
    \label{fig:param_choice}
\end{figure}

Let us begin by considering a general density observable,
\begin{equation}
\rho(\vec{r}) = \sum_{m,n} \rho_{mn}\omega^{mr_1+nr_2},
\end{equation}
where $\vec{r}=r_1\vec{a}_1 + r_2\vec{a}_2$ ($\vec{a}_{1,2}\equiv$basis vectors) and the $\rho_{mn}$ are order parameters for the various density wave states with wave vector $\vec{Q}_{mn} = (m\vec{b}_1 + n\vec{b}_2)/3$, and $\vec{b}_1,\vec{b}_2$ are shown in Fig.~\ref{fig:param_choice}. In the $N=1$ limit of Eq.~(\ref{eqn:large_N}), the $\rho_{mn}$ can be expressed in terms of the vortex flavors as \cite{burkov_superfluid-insulator_2005} 
\begin{equation}
\rho_{mn} \propto \omega^{-mn/2+(n-m)/6}\sum_{l} \phi^*_l \phi_{l+m+2n}\omega^{m(l+m+n-1)} \label{eqn:rho_mn},
\end{equation}
with a proportionality scalar factor, $S(m,n)$, not prescribed solely by the PSG. When we promote $\phi_l$ to $\vec{\phi}_l$ we will require that the global $SU(N)$ symmetry of our theory space is unbroken, so the $\phi_l$ variables in the above relation will be trivially extended to $\vec{\phi}_l$.

We first consider the relevance of the order parameters, $\rho_{mn}$, for $m\neq n$ . These can include the density wave states with the ordering wavevectors seen in Figure~\ref{fig:RG_flow}(a), i.e., those we expect at the mean-field level. Based on Eqn.~\ref{eqn:rho_mn}, we conclude that $\rho_{mn}$ with $m\neq n$ does not transform as a singlet operator; the na\"{i}ve scaling dimension $\Delta_{\rho_{mn}} = (d-2+\eta_{\rho_{mn}})/2 = 1$, and thus $\eta_{\rho_{mn}} = 1$. Based on our earlier arguments of Sec.~\ref{sec:criteria}, we conclude that this is a marginal coupling and so the effect of the charge order on the spinon Fermi surface must be considered. However, the ordering wave vector, $\vec{Q}$, will only couple the spinon Fermi surface to the chargon density-wave if $\vec{Q}$ connects points along the Fermi surface that are separated by $2k_F$, leading to ``hot spots." Fortunately, at the densities of interest and for sufficiently generic Fermi surfaces, such hot spots can be avoided altogether (i.e., $|2k_F|<Q$). Consider, for instance, a dispersion generated by including up to third-nearest-neighbor hopping on the triangular lattice, with $t_2/t_1 = 2, t_3/t_1 = -1.25$ (Fig.~\ref{fig:param_choice}); we can avoid the hot spots altogether for $\vec{Q} = \vec{b}_2/3$. In fact this choice of hoppings will avoid hot spots for all $\vec{Q}_{mn}, m\neq n$. This serves as a proof of principle that it is not unreasonable for the hot spots to be absent for  $\vec{Q}_{mn}, ~m\neq n$, at the densities of interest. 

While the Fermi surface may avoid hot spots at the wave vectors noted above, it exhibits near nesting for $\vec{Q}_{11}$ and $\vec{Q}_{22}$. These correspond to the $\sqrt{3}\times\sqrt{3}$ charge order discussed earlier (Fig.~\ref{fig:phase_transition}), and we thus must consider the relevance of this charge order. We note that for $\rho_{mn}$ with $m=n$, $\rho_{nn} \propto \sum_l |\bs{\phi}_l|^2 \omega^{nl}$. This is a linear combination of singlet operators at the decoupled $O(2N)$ fixed point. Hence, $\Delta_{\rho_{nn}} = \Delta_{S} = 2$, leading to $\eta_{\rho_{nn}} = 3$. Using the same arguments from earlier, $\rho_{nn}$ is now irrelevant at the fixed point. Thus, even when the Fermi surface is nearly nested for these special wave vectors and can host hot spots (Fig.~\ref{fig:param_choice}), any possible $\sqrt{3}\times\sqrt{3}$ density-wave in the chargon sector will decouple dynamically from the spinon Fermi surface.

Finally, we must also examine the effects of coupling the particle-hole fluctuations near the Fermi surface to the nematic order $\N$, as outlined in Eq.~(\ref{Scdw}). The lowest order possible nematic parameter constructed out of the vortex degrees of freedom that preserves the global $SU(N)$ symmetry will be of the form
\begin{equation}
\N(\bs{\phi},\bs{\phi}^\dagger) = \sum_{l,k} c_{lk} \bs{\phi}_l^\dagger \bs{\phi}_k,
\end{equation}
where $c_{lk}\in \mathbb{C}$. In the presence of only a spontaneously broken rotational symmetry (i.e., no other form of broken translational symmetry), we have $c_{lk} = c_{l-1,k-1}$ and $c_{lk} = \omega^{l-k}c_{l-1,k-1}$ from $T_{1}$ and $T_{2}$ in Table \ref{tab:PSG}. But then we must have that $c_{lk} \propto \delta_{lk}$ and $\N \propto |\bs{\phi}|^2$. This is clearly invariant under $R_{2\pi/3}$ and thus cannot be a valid nematic order parameter. Thus $\N$ must be at least quartic in the $\phi$ fields, since it must also preserve global $U(1)$ symmetry. This means $\Delta_{\N} = (d - 2 + \eta_{\N})/2 \geq 2$ and thus $\eta_{\N} \geq 3$. Once again, the coupling of the nematic order to the spinon Fermi surface is irrelevant.

To conclude, for our specific large$-N$ generalization of the model, we arrive at the remarkable result that the dual vortex theory for the chargons decouples dynamically from the spinons for $N\rightarrow\infty$. In our discussion so far,  we did not explicitly state the role of the transverse gauge-field fluctuations, $a$. Including the effects of these fluctuations does not modify these conclusions and the matter fields remain decoupled (see Appendix \ref{app:phys_quant_QCP}). The situation is reminiscent of the bandwidth-tuned transition at $\nu_c=1/2$ \cite{senthil_theory_2008}; the Landau damping term for $a$ that is generated from the Fermi surface behaves like a ``Higgs mass" term when the dynamical critical exponent $z=1$ for the chargons. Thus, the fluctuations of $a$ do not affect the chargon dynamics, and neither can the $f$ affect the chargons indirectly via a coupling to $a$.

\section{Experimental signatures}
\label{sec:expt}

Based on our theoretical framework, we can make a number of predictions for experimentally measurable quantities near the metal-WM insulator critical point. Interestingly, a number of these signatures are {\it qualitatively} similar to the behavior near the bandwidth-tuned transition for $\nu_c=1/2$ \cite{senthil_theory_2008}, while the exact critical singularities are different. The technical details are summarized in Appendix \ref{app:phys_quant} and we only focus on the results here.

As the critical point is approached from the insulating side the charge gap will vanish continuously with exponent $z\nu$, where $z=1$; within our large-$N$ formulation, $\nu z=1$. Going beyond this limit, even to $N=1$, a second-order transition of the kind we describe is only possible if $\nu > 2/3, z=1$.

Upon approaching the critical point from the metallic side, the fermion self energy due to scattering off the fluctuations of the renormalized $U(1)$ gauge field will be given by
\begin{equation}
\Sigma_f(\vec{K},i\omega) = ia\omega \begin{cases} 2\ln(1/\rho_s) &\mbox{on the FL side}\\ \ln(1/|\omega|) &\mbox{at the critical point}\end{cases},
\end{equation}
where $a$ is a constant and $\vec{K}$ lies on the spinon Fermi surface. At the critical point we recover a marginal Fermi liquid form for the self-energy; in the Fermi liquid (FL), the singular form is cut off by $\rho_s \sim 1/\xi$, the superfluid stiffness associated with the chargons, and $\xi$ the correlation length. On the FL side, the quasiparticle residue will behave as $Z \sim |s-s_c|^{2\beta}/\ln(1/|s-s_c|)$, where $s$ is the parameter tuning the transition. For the theory at large$-N$, we have $\beta=1/2$. Additionally the effective mass of the quasiparticle will diverge logarithmically as the transition is approached from the FL side, which will likely manifest itself in a Kadowaki-Woods scaling of the coefficient of the $T^2-$resistivity and in the Sommerfeld coefficient associated with the specific heat.
The Ioffe-Larkin rule (Appendix \ref{app:Ioffe_Larkin}) for the electronic compressibility, $\kappa^{-1} = \kappa_b^{-1} + \kappa^{-1}_f$, with $\kappa_b\sim \rho_s$, also leads to the conclusion that $\kappa$ vanishes continuously upon approaching the critical point from the metallic side.

The Ioffe-Larkin rules also determine the evolution of the resistivity across the critical point. In the clean limit, there is a jump in the resistivity at the critical point of order $\sim Rh/e^2$, with $R$ a universal number \cite{senthil_theory_2008, witczak-krempa_universal_2012}. However, there is a minor modification beyond the considerations of the bandwidth-tuned transition at $\nu_c=1/2$, arising from the structure of the dual vortex multiplet theory. Specifically, the fixed point in the large$-N$ limit of our theory is described by three $\mathbb{C}\mathbb{P}^{N-1}$ models coupled by the noncompact $U(1)$ gauge field $A_\mu$. The WM insulator corresponds to each of the $\bs{\phi}_l$ condensed, where each $\mathbb{C}\mathbb{P}^{N-1}$ will have a topological defect that corresponds to the winding of the phase of $\bs{\phi}_l$. As noted earlier, these defects correspond to fractionalized chargons and carry charge $e/3$ \cite{burkov_superfluid-insulator_2005} and the universal jump in resistivity at the critical point will be given by
\begin{equation}
\rho_b = \frac{Rh}{3(e/3)^2} = \frac{3Rh}{e^2},
\end{equation}
which is three times the size of the jump absent any fractionalization. This enhancement is discussed in greater detail in \cite{xu_metal-insulator_2021}. Note that the above relation assumes the underlying large-$N$ limit. More generally, the universal jump in the resistivity at this filling will have a magnitude that is different from the corresponding jump at half filling.

Turning our attention now to the effects of translation symmetry breaking, which is a new ingredient beyond the MIT at $\nu_c=1/2$, there is a continuous onset of charge density order upon entering the WM insulator. The density-density correlations for the various order parameters are controlled by the corresponding scaling dimensions, $\Delta_{\rho}$. Specifically for our large$-N$ formulation of the theory, we anticipate the $\sqrt{3}\times \sqrt{3}$ ordering to have a scaling dimension $\Delta_{\rho_{nn}} = 2$, with all other charge orderings having scaling dimension $1$.

The magnetic properties across the MIT can be probed through measurements of the spin susceptibility, as has already been done in the recent experiments of $\nu_c=1/2$ bandwidth-tuned transition in moir\'e TMD materials \cite{li_continuous_2021}. For the MIT discussed in this paper, the spin susceptibility in the WM will be temperature independent at the lowest temperatures due to the presence of the spinon Fermi surface. Moreover, the susceptibility will evolve smoothly across the transition to the expected Fermi liquid form on the metallic side. However, the bandwidth associated with the spinon Fermi surface is expected to be small, as it is controlled by the exchange interactions that are generated due to the longer range hopping and interaction scales in the ordered Wigner crystal. Therefore, as a function of increasing temperatures, the spinons are expected to crossover into their `high-temperature' state and can exhibit an analog of a `Pomeranchuk-effect' with a significant enhancement in their entropy from the fluctuating local moments. 

\section{Outlook}
\label{sec:outlook}

Preliminary thermodynamic measurements studying the bandwidth-tuned transition at fixed commensurate fillings out of certain WM insulators in moir\'e TMD materials find no evidence of hysteresis \cite{li_continuous_2021}. These platforms provide an ideal playground for investigating the continuous metal-insulator transitions that have been the focus of the present theoretical study. We have used a large-$N$ approach to describe the transition analytically in a controlled fashion and demonstrated the possibility of realizing such a continuous transition without any fine-tuning. However, the physical situation at $N=1$ lies beyond the strict regime of control within our approach and is possibly described by an entirely different fixed point. Moreover, the specific form of translational symmetry breaking at large$-N$ in our theory is different from the $\sqrt{3}\times \sqrt{3}$ charge ordering observed experimentally \cite{li_imaging_2021}. It would be very interesting and helpful to the community to use a combination of more sophisticated numerical and analytical techniques to directly study the strongly coupled theory in the $N=1$ limit in the future.

The Wigner-Mott insulator in the present theoretical discussion hosts a Fermi surface of neutral spinons. It is natural to address the possibility of a direct transition from a metal to a Wigner-Mott insulator {\it without} gapless Fermi surfaces of {\it any} excitations. Such a transition presents a significant challenge to theory and requires the spinon Fermi surface to disappear infinitesimally away from the critical point on the insulating side without any fine-tuning. A possible route to describing such transitions has been discussed in another context recently \cite{zou_deconfined_2020,YZSS20,mandal_critical_2020}.

Finally, it would be interesting to study the effect of doping slightly away from $\nu_c=1/6$ on the insulating side. There are at least three possibilities. In the simplest scenario, the excess carriers form a small Fermi surface, behaving as a spectator to the MIT described above without altering its criticality. However, at these low densities in the experimental setup, quantum localization corrections and effects of long-wavelength disorder likely play an important role. More interestingly, if the excess carriers get fractionalized due to strong-correlation effects on the insulating side, the corresponding chargons can condense. However, this can immediately lead to a transition into a metallic state where the spinon Fermi surfaces reveal themselves and become electronic Fermi surfaces consistent with the full Luttinger count in the reduced Brillouin zone (due to the underlying translational symmetry breaking). Additionally, we also note that the excess doping could, in principle, enlarge the spinon Fermi surface enough to introduce ``hot spots" at the relevant charge-ordering wave vectors, and change the nature of the underlying fixed point.

\acknowledgements

We thank K. F. Mak, J. Shan, and W. Zhao for numerous insightful discussions. S.M. is supported by the National Science Foundation Graduate Research Fellowship under Grant No. 1745302. Any opinions, findings, and conclusions or recommendations expressed in this material are those of the author(s) and do not necessarily reflect the views of the National Science Foundation. S.M. would like to thank LASSP at Cornell University, where a portion of this work was completed, for their hospitality. T.S. was supported by U.S. Department of Energy Grant No. DE- SC0008739, and partially through a Simons Investigator Award from the Simons Foundation. This work was also partly supported by the Simons Collaboration on Ultra-Quantum Matter, which is a grant from the Simons Foundation (Grant No. 651440, T.S.). D.C. is partially supported by Grant No. 2020213 from the United States-Israel Binational Science Foundation (BSF), Jerusalem, Israel.  

\appendix

\section{Mean-field theory of electronic density-wave ordering}

\subsection{$120^{\circ}$ N\'{e}el order}
\label{app:HF_120}
Consider a single spin density wave of the form discussed in Sec.~\ref{sec:prelimHF}, e.g. one that rotates the spins by $120^{\circ}$ as they increment along the $\vec{a}_2$ direction (shown in Fig.~\ref{fig:d_lat}) but leaves them unchanged in the $\vec{a}_1$ direction. The spin wave order parameter will thus be given by 
\begin{align}
\vec{m}_i =& \Delta_{S_2}\left[\cos\left(\frac{1}{3} (\vec{b}_2 \cdot \vec{r}_i)\right)\vec{\hat{x}}+\sin\left(\frac{1}{3} (\vec{b}_2 \cdot \vec{r}_i)\right)\vec{\hat{y}}\right]\\
\vec{m}_{\vec{q}} =& \frac{\Delta_{S_2}}{2}\left[\left(\delta_{\vec{q},\vec{Q}_2} + \delta_{\vec{q},-\vec{Q}_2}\right)\vec{\hat{x}} - i\left(\delta_{\vec{q},\vec{Q}_2} - \delta_{\vec{q},-\vec{Q}_2}\right)\vec{\hat{y}}\right],
\end{align}
where $\vec{Q}_2 = \vec{b}_2/3$ is the spin density wave vector. The spin density wave (SDW) Hamiltonian will then be 
\begin{align}
H_{\mathrm{SDW},2} =& \sum_{\vec{k},\vec{q}} c^\dagger_{\vec{k},\alpha} \vec{m}_{\vec{q}}\cdot \bs{\sigma}_{\alpha\beta} c_{\vec{k}+\vec{q},\beta}\\
=& \Delta_{S_2}\sum_{\vec{k}} c^\dagger_{\vec{k},\uparrow}c_{\vec{k}+\vec{Q}_2,\downarrow} + c^\dagger_{\vec{k},\downarrow}c_{\vec{k}-\vec{Q}_2,\uparrow}\label{eqn:HSDW2}.
\end{align}
As discussed in Sec.~\ref{sec:prelimHF} the full SDW will be described by competing SDW states: one state that wants the spins to increment by $120^{\circ}$ as they travel along $\vec{a}_2$ but be aligned along $\vec{a}_1$ and the other which wants alignment along $\vec{a}_2$. This second state will be given by $H_{\mathrm{SDW},1}$, otherwise written the same as above but with the order parameter $\Delta_{S_1}$ and the spin density wave vector now given by $\vec{Q}_1 = \vec{b}_1/3$. The full Hamiltonian will then be given by
\begin{equation}
H = H_t + H_{\mathrm{SDW},1} + H_{\mathrm{SDW},2},
\end{equation}
where $H_t$ is the hopping Hamiltonian. Since $\vec{Q}_1$ and $\vec{Q}_2$ are both commensurate with the original Brillouin zone the bands can be folded into the reduced Brillouin zone spanned by $\vec{Q}_1$ and $\vec{Q}_2$. As each $\vec{Q}$ has periodicity $3$ the enlarged unit cell will contain $9$ sites, as we noted it must. Accounting for spin this means there will be $18$ bands. This gives us a chance of observing a band gap at fixed filling $\nu_c=1/6$.

We can numerically tune the order parameters $\Delta_{S_i}$ with a specific model of the hopping and confirm that there is indeed a band gap at $\nu_c = 1/6$ filling at large $\Delta_S$. First, we take $\Delta_{S_1} = \Delta_{S_2} = \Delta$ in order to get maximal frustration and encourage the charge density wave pattern seen. Next we consider
\begin{align}
H_t = \sum_{\vec{k}} \xi_{\vec{k}} c^\dagger_{\vec{k}\sigma}c_{\vec{k}\sigma} \label{eqn:H_t},
\end{align}
where $\xi_{\vec{k}}$ is the dispersion given by nearest-neighbor hopping on the triangular lattice, accompanied by a chemical potential term meant to keep the system at $\nu_c=1/6$. The results are displayed as Fig.~\ref{fig:HF}(a). With increasing $\Delta/t$ the Fermi surface is eventually gapped out. We further measure this gap to be proportional to $\Delta$, so it is not sensitive to the details of the hopping Hamiltonian.

\subsection{Stripe order}
\label{app:HF_stripe}
In the case of the stripe order shown in Fig.~\ref{fig:HF}(b) we have a SDW with period $2$ along the $\vec{a}_2$ direction and a charge density wave with period $3$ along the $\vec{a}_1$ direction. The SDW Hamiltonian is given by
\begin{equation}
H_{\mathrm{SDW}} = \Delta_S\sum_{\vec{k}} c^\dagger_{\vec{k},\alpha} \sigma^z_{\alpha\beta} c_{\vec{k}+\vec{Q}_2,\beta},
\end{equation}
where $\vec{Q}_2 = \vec{b}_2/2$. The charge density wave order will not couple to the spins and can be written simply as
\begin{align}
H_{\mathrm{CDW}} = \Delta_C\sum_{\vec{k},\sigma} c^\dagger_{\vec{k},\sigma}c_{\vec{k}+\vec{Q}_1,\sigma},
\end{align}
where $\vec{Q}_1 = \vec{b}_1/3$. Note that in real space this is given by
\begin{equation}
H_{\mathrm{CDW}} = \sum_{\vec{r}_i,\sigma} \left[\Delta_C\cos\left(\frac{1}{3}(\vec{b}_1\cdot \vec{r}_i)\right)\right]n_{\vec{r}_i,\sigma}.
\end{equation}
Since we want to make it energetically favorable to fill the first site, $\vec{r}_i=0$, and not the next two, $\vec{r}_i = \vec{a}_1,2\vec{a}_1$, we must take $\Delta_C < 0$. Finally, we note that the enlarged unit cell will now contain $6$ original lattice sites, making it possible to observe a band insulator with filling $\nu_c=1/6$.

In order to numerically check this we take $-\Delta_C = \Delta_S = \Delta >0$. With the same hopping Hamiltonian Eq.~(\ref{eqn:H_t}) we find that a gap does open up for $\nu_c=1/6$ as $\Delta/t$ is increased. Further we find that this gap is proportional to $\Delta$ for large $\Delta$, so it is again not sensitive to the details of the hopping Hamiltonian.

\section{Further details of the low-energy field theory}
\label{app:phys_quant}

In this appendix, we will compute the effective action of the emergent gauge field $a_\mu$ and use this to understand the physics of the low-energy field theory.

\subsection{Effective action for emergent $U(1)$ gauge-field}
\label{app:Seff_a_FL}

The effective action for the emergent gauge field, $a_\mu$, includes dynamical contributions from both the gapless spinons and the chargon fields. The spinons couple minimally to $a$ and lead to a familiar contribution, once the particle-hole fluctuations near the Fermi surface are integrated out. At small $|\omega|$ and $|\vec{q}|$ (the momentum deviation from the FS), the spinon polarizability is
\begin{equation}
\Pi_f(\vec{q},i\omega) = \frac{k_0|\omega|}{v_{F_0}|\vec{q}|} + \chi_d |\vec{q}|^2 + \cdots, \label{eqn:Pi_af}
\end{equation}
where $k_0$ is of order the typical spinon Fermi momentum, $v_{F_0}$ is the typical spinon Fermi velocity, and $\chi_d$ represents the diamagnetic susceptibility of the spinons.

In the original chargon description, the emergent gauge field $a_\mu$ will couple via a term $a_\mu j^\mu_b$, where $j^\mu_b$ is the chargon number current. On the dual side, this current is given by $\epsilon^{\mu\nu \lambda} \partial_\nu A_\lambda/2\pi$; physically this is because the presence of a boson corresponds to $2\pi$ flux from $A$. Thus on the dual side there will be no direct coupling of $a_\mu$ to the vortices, instead they will be coupled indirectly via a mutual Chern-Simons term with $A_\mu$. The effective action for the transverse parts of $A$ and $a$ will then be given by
\begin{align}
\S_{\mathrm{eff}}[a, A] =& \int_{\vec{q},\omega} \bigg[\Pi_{[a,f]}(\vec{q},i\omega)|a(\vec{q},\omega)|^2 + \frac{i \epsilon^{\mu \nu\lambda}q_\nu}{2\pi}A_\mu a_\lambda\bigg] \nonumber\\
&+ \S_{\mathrm{eff}}[A], \text{ where} \label{eqn:Seff_aA}\\
\Pi_{[a,f]}(\vec{q},i\omega) =& \frac{q^2}{e^2_a} + \Pi_f(\vec{q},i\omega)
\end{align}
and $\S_{\mathrm{eff}}[A]$ is the effective quadratic action in terms of $A_\mu$ when the vortex fields are integrated out, which we return to later. Note the presence of an $i$ factor in the mutual Chern-Simons term above; this is because of its extra factor of $i$ even in a Euclidean path integral framework.
The effective action for the transverse components of $a$, after integrating out the $A$ field, is given by
\begin{equation}
\S_{\mathrm{eff}}[a] = \int_{\vec{q},\omega} \left(\Pi_{[a,f]}(\vec{q},i\omega) + \frac{q^2}{(4\pi)^2 \Pi_A(\vec{q},i\omega)}\right)|a(\vec{q},\omega)|^2, \label{eqn:Seff_a}
\end{equation}
where $\Pi_A(\vec{q},i\omega)$ is the vortex polarizability. We thus see that the behavior of the effective action for $a_\mu$ when the matter fields are integrated out requires knowledge of $\Pi_A$. In order to calculate this in a sensible way the $N$ counting must be done correctly. With the action given in Eq.~(\ref{eqn:large_N}) the gauge field $A_\mu$ will drop out of any correlators involving the vortex multiplets in the $N\rightarrow \infty$ limit \cite{kaul_quantum_2008}. However, the reverse is not true and the $\phi_l$ fields will affect the effective action of $A$ when they are integrated out, i.e., make $\Pi_A$ nontrivial.

\subsection{Approach from Fermi liquid side}

After integrating out the gapped vortex fields (i.e., condensed chargons) within our large$-N$ formulation, the effective action for (the transverse part of) $A_\mu$ to quadratic order is given by
\begin{equation}
\S_{\mathrm{eff}}[A] = \int_{\vec{q},\omega} \left(\frac{q^2}{e^{ 2}_A}+\Pi_\phi(\vec{q},i\omega)\right)|A(\vec{q},\omega)|^2,
\end{equation}
where $\Pi_\phi$ is the polarizability of a \textit{single} species of vortices; the $N-$normalization is chosen such that $\Pi_\phi$ is $\mathcal{O}(1)$. The polarizability will have the form
\begin{equation}
\Pi_\phi(\vec{q},i\omega) = \sigma_\phi \sqrt{\omega^2 + c|\vec{q}|^2} P\left(\frac{\sqrt{\omega^2 + c|\vec{q}|^2}}{\rho_s}\right),
\end{equation}
where $P$ is some function which scales as $P(x\rightarrow 0)\sim x/\pi$ and $P(x\rightarrow \infty)\sim 1$, $\sigma_\phi$ is the universal vortex conductivity at the critical point, and $\rho_s$ and $c$ are the superfluid stiffness and velocity of the chargons, respectively.

We then obtain
\begin{equation}
\Pi_A(\vec{q}, i\omega) = \frac{q^2}{e^{ 2}_A} + \sigma_\phi q P\left(\frac{q}{\rho_s}\right) \label{eqn:Seff_A},
\end{equation}
where $q = \sqrt{\omega^2 + c|\vec{q}|^2}$. In the Fermi liquid, $\rho_s>0$, such that for $q\ll \rho_s$, $P(q/\rho_s)\sim q/(\pi\rho_s)$. As expected, integrating out the gapped $\phi$ fields leads to nonsingular terms that simply renormalize $1/e^{ 2}_A \rightarrow 1/e^{ 2}_A + \sigma_\phi/\pi \rho_s$. The effective theory for $a$ at low energies is then given by
\begin{align}
S_{\mathrm{eff}}[a] =& \int_{\vec{q},\omega} \left(\frac{k_0|\omega|}{v_{F_0}|\vec{q}|} +  \frac{q^2}{(4\pi)^2[q^2/e^{ 2}_A + \sigma_\phi q^2/\pi \rho_s]}\right) |a(\vec{q},\omega)|^2 \nonumber\\
\sim& \int_{\vec{q},\omega} \left(\frac{k_0|\omega|}{v_{F_0}|\vec{q}|} + \frac{\pi \rho_s}{(4\pi)^2 \sigma_\phi}\right)|a(\vec{q},\omega)|^2, 
\end{align}
as $\rho_s$ becomes small relative to $e_A$. As in the example of the MIT at $\nu_c=1/2$ \cite{senthil_theory_2008}, the term proportional to $\rho_s$ cuts off the singular divergence of the fermion self-energy due to scattering off the $a-$fluctuations, leading to a regular $\Sigma_f(\vec{K},i\omega) \sim i\omega \ln(1/\rho_s)$. The $\rho_s$ term is an analog of a Higgs term for the emergent gauge fields at low energies. This term is proportional to the chargon phase stiffness $\rho_s$. Clearly, at the critical point $\rho_s\rightarrow 0$ and the critical vortex (or, equivalently, the chargon) fluctuations will lead to singular contributions to the effective action for $a$, which will affect the fermion self-energy. 

\subsection{Low energy theory at the critical point}
\label{app:phys_quant_QCP}

At the critical point, the form of the effective action for $a$ is modified primarily due to the contribution from the gapless bosonic matter fields. In this limit, the action becomes
\begin{align}
S_{\mathrm{eff}}[a] =& \int_{\vec{q},\omega} \left(\frac{k_0|\omega|}{v_{F_0}|\vec{q}|} + \frac{c^2|\vec{q}|^2}{(4\pi)^2[c^2|\vec{q}|^2/e^{ 2}_A + \sigma_\phi c|\vec{q}|]}\right) |a|^2 \nonumber\\
=& \int_{\vec{q},\omega} \left(\frac{k_0|\omega|}{v_{F_0}|\vec{q}|} + \frac{c}{(4\pi)^2 \sigma_\phi}|\vec{q}|\right) |a(\vec{q},\omega)|^2 \label{eqn:Seff_a_low_energy}.
\end{align}

We note that the above low-energy effective action is nearly identical to the critical theory for the bandwidth-tuned transition at $\nu_c=1/2$ \cite{senthil_theory_2008}. As expected, our formulation leads to the appearance of the vortex resistivity, $\sim1/\sigma_\phi$, instead of the boson conductivity, $\sigma_0$, which are related to each other \cite{fisher_quantum_1990}. On a quantitative level, this universal conductivity (resistivity) describes the contribution from a superfluid to charge-order deconfined critical point, unlike that of the 3D XY critical point for the $\nu_c=1/2$ transition.

As anticipated, the effect of the above $z=2$ gauge field is to lead to a $\Sigma_f(\vec{K},i\omega) \sim i \omega \ln(1/|\omega|)$ form of the fermion self-energy at the critical point; the chargon self-energy receives only analytic corrections from the gauge field. The chargons are described by $z_b=1$ and the Landau damping term in Eq.~(\ref{eqn:Seff_a_low_energy}) effectively behaves as a Higgs mass. As a result, the fluctuations of $a$ in the chargon sector and their criticality will thus be unaffected by $a$, and hence by any indirect coupling from $f$ to $a$.

\subsection{Ioffe-Larkin rules}
\label{app:Ioffe_Larkin}

In this appendix, we review the Ioffe-Larkin rules for the effective electromagnetic (and related) response functions, starting from the dual vortex side. We imagine coupling the system to an external (probe) $U(1)$ gauge field $A^{\mathrm{ext}}_\mu$. Since we have assigned the physical electric charge to the chargons, and not the spinons, $A^{\mathrm{ext}}_\mu$ only couples to these matter fields. In the original chargon description this coupling would be of the form $A_\mu^{\mathrm{ext}}j^\mu_b$ with $j^\mu_b$ the chargon number current. By the same arguments given in Appendix \ref{app:Seff_a_FL}, we see that on the dual vortex side $A^{\mathrm{ext}}$ will thus couple to $\epsilon^{\mu\nu\lambda}\partial_\nu A_\lambda/2\pi$ and the effective action in terms of $a$ and $A$ [Eq.~(\ref{eqn:Seff_aA})] will gain a term
\begin{equation}
-\int_{\vec{q},\omega} \frac{i}{2\pi}A_\mu \epsilon^{\mu\nu\lambda}q_\nu A^{\mathrm{ext}}_{\lambda}.
\end{equation}
When $A$ is integrated out, as done in Appendix \ref{app:Seff_a_FL}, we will now have that the effective action in terms of $a$ is given by
\begin{align}
\S_{\mathrm{eff}}[a,A^{\mathrm{ext}}] =& \int_{\vec{q},\omega}\left[\Pi_{f}(\vec{q},i\omega)|a|^2 + \Pi_b(\vec{q},i\omega)|a-A^{\mathrm{ext}}|^2\right],\\
\Pi_b(\vec{q},i\omega) =& \frac{q^2}{(4\pi)^2 \Pi_A(\vec{q},i\omega)}.
\end{align}
The coefficient of the $a-A^{\mathrm{ext}}$ term is $\Pi_b(\vec{q},i\omega)$, since this is how it must appear in the chargonic description of the theory. We have also dropped the unrenormalized $a$ propagator from the $\Pi_{[a,f]}$ term since this will turn out to be irrelevant for our interests. If we now integrate out the internal gauge field $a_\mu$ we recover the Ioffe-Larkin result \cite{ioffe_gapless_1989} for the effective theory in terms of the external gauge field $A^{\mathrm{ext}}$,
\begin{equation}
\Pi(\vec{q},i\omega)^{-1} = \Pi_f(\vec{q},i\omega)^{-1} + \Pi_b(\vec{q},i\omega)^{-1} \label{eqn:Ioffe_Larkin}.
\end{equation}

The above can be used immediately to obtain, e.g., the compressibility and resistivity. In terms of the polarizability the compressibility of a system is given by
\begin{equation}
\kappa = \lim_{|\vec{q}|\rightarrow 0} \Pi(\vec{q},i\omega=0),
\end{equation}
such that $\kappa^{-1} = \kappa^{-1}_b + \kappa^{-1}_f$. On the FL side where $\rho_s>0$, we have
\begin{equation}
\kappa_b = \frac{\rho_s}{16\pi[\pi \rho_s/e^{ 2}_A + \sigma_\phi]} \sim \frac{\rho_s}{16\pi \sigma_\phi}.
\end{equation}
Thus $\kappa_b \sim \rho_s \sim 1/\xi$ as the critical point is approached.

The conductivity is given by
\begin{equation}
\sigma = \lim_{\omega \rightarrow 0} \frac{1}{\omega}\mathrm{Im}\Pi(\vec{q}=0, i\omega \rightarrow \omega+i\delta),
\end{equation}
and $\rho = \rho_b + \rho_f$. In the presence of some weak disorder, $\sigma_f$ is finite; the chargon conductivity is infinite on the FL side, while on the insulating side it must be zero. At the critical point we will have
\begin{equation}
\sigma_b = \lim_{\omega\rightarrow 0} \frac{1}{\omega}\mathrm{Im}\left[\frac{-\omega^2}{(4\pi)^2[-\omega^2/e^{ 2}_A + i\sigma_\phi \omega]}\right] = \frac{1}{(4\pi)^2\sigma_\phi}.
\end{equation}
Note that the universal value at the critical point for $\sigma_\phi$ will determine the ``jump" in the resistivity (the `$Rh/e^2$' coefficient introduced earlier).  

\bibliographystyle{apsrev4-1_custom}
\bibliography{MIT.bib}

\begin{thebibliography}{34}%
\makeatletter
\providecommand \@ifxundefined [1]{%
 \@ifx{#1\undefined}
}%
\providecommand \@ifnum [1]{%
 \ifnum #1\expandafter \@firstoftwo
 \else \expandafter \@secondoftwo
 \fi
}%
\providecommand \@ifx [1]{%
 \ifx #1\expandafter \@firstoftwo
 \else \expandafter \@secondoftwo
 \fi
}%
\providecommand \natexlab [1]{#1}%
\providecommand \enquote  [1]{``#1''}%
\providecommand \bibnamefont  [1]{#1}%
\providecommand \bibfnamefont [1]{#1}%
\providecommand \citenamefont [1]{#1}%
\providecommand \href@noop [0]{\@secondoftwo}%
\providecommand \href [0]{\begingroup \@sanitize@url \@href}%
\providecommand \@href[1]{\@@startlink{#1}\@@href}%
\providecommand \@@href[1]{\endgroup#1\@@endlink}%
\providecommand \@sanitize@url [0]{\catcode `\\12\catcode `\$12\catcode
  `\&12\catcode `\#12\catcode `\^12\catcode `\_12\catcode `\%12\relax}%
\providecommand \@@startlink[1]{}%
\providecommand \@@endlink[0]{}%
\providecommand \url  [0]{\begingroup\@sanitize@url \@url }%
\providecommand \@url [1]{\endgroup\@href {#1}{\urlprefix }}%
\providecommand \urlprefix  [0]{URL }%
\providecommand \Eprint [0]{\href }%
\providecommand \doibase [0]{http://dx.doi.org/}%
\providecommand \selectlanguage [0]{\@gobble}%
\providecommand \bibinfo  [0]{\@secondoftwo}%
\providecommand \bibfield  [0]{\@secondoftwo}%
\providecommand \translation [1]{[#1]}%
\providecommand \BibitemOpen [0]{}%
\providecommand \bibitemStop [0]{}%
\providecommand \bibitemNoStop [0]{.\EOS\space}%
\providecommand \EOS [0]{\spacefactor3000\relax}%
\providecommand \BibitemShut  [1]{\csname bibitem#1\endcsname}%
\let\auto@bib@innerbib\@empty
\bibitem [{\citenamefont {Senthil}(2008{\natexlab{a}})}]{senthil2008critical}%
  \BibitemOpen
  \bibfield  {author} {\bibinfo {author} {\bibfnamefont {T.}~\bibnamefont
  {Senthil}},\ }\bibfield  {title} {\enquote {\bibinfo {title} {Critical fermi
  surfaces and non-fermi liquid metals},}\ }\href@noop {} {\bibfield  {journal}
  {\bibinfo  {journal} {Physical Review B}\ }\textbf {\bibinfo {volume} {78}},\
  \bibinfo {pages} {035103} (\bibinfo {year} {2008}{\natexlab{a}})}\BibitemShut
  {NoStop}%
\bibitem [{\citenamefont {Senthil}(2008{\natexlab{b}})}]{senthil_theory_2008}%
  \BibitemOpen
  \bibfield  {author} {\bibinfo {author} {\bibfnamefont {T.}~\bibnamefont
  {Senthil}},\ }\bibfield  {title} {\enquote {\bibinfo {title} {Theory of a
  continuous {Mott} transition in two dimensions},}\ }\href {\doibase
  10.1103/PhysRevB.78.045109} {\bibfield  {journal} {\bibinfo  {journal}
  {Physical Review B}\ }\textbf {\bibinfo {volume} {78}},\ \bibinfo {pages}
  {045109} (\bibinfo {year} {2008}{\natexlab{b}})}\BibitemShut {NoStop}%
\bibitem [{\citenamefont {Padhi}\ \emph {et~al.}(2021)\citenamefont {Padhi},
  \citenamefont {Chitra},\ and\ \citenamefont
  {Phillips}}]{padhi_generalized_2021}%
  \BibitemOpen
  \bibfield  {author} {\bibinfo {author} {\bibfnamefont {B.}~\bibnamefont
  {Padhi}}, \bibinfo {author} {\bibfnamefont {R.}~\bibnamefont {Chitra}}, \
  and\ \bibinfo {author} {\bibfnamefont {P.~W.}\ \bibnamefont {Phillips}},\
  }\bibfield  {title} {\enquote {\bibinfo {title} {Generalized {Wigner}
  crystallization in moiré materials},}\ }\href {\doibase
  10.1103/PhysRevB.103.125146} {\bibfield  {journal} {\bibinfo  {journal}
  {Physical Review B}\ }\textbf {\bibinfo {volume} {103}},\ \bibinfo {pages}
  {125146} (\bibinfo {year} {2021})}\BibitemShut {NoStop}%
\bibitem [{\citenamefont {Tang}\ \emph {et~al.}(2020)\citenamefont {Tang},
  \citenamefont {Li}, \citenamefont {Li}, \citenamefont {Xu}, \citenamefont
  {Liu}, \citenamefont {Barmak}, \citenamefont {Watanabe}, \citenamefont
  {Taniguchi}, \citenamefont {MacDonald}, \citenamefont {Shan},\ and\
  \citenamefont {Mak}}]{tang_simulation_2020}%
  \BibitemOpen
  \bibfield  {author} {\bibinfo {author} {\bibfnamefont {Y.}~\bibnamefont
  {Tang}}, \bibinfo {author} {\bibfnamefont {L.}~\bibnamefont {Li}}, \bibinfo
  {author} {\bibfnamefont {T.}~\bibnamefont {Li}}, \bibinfo {author}
  {\bibfnamefont {Y.}~\bibnamefont {Xu}}, \bibinfo {author} {\bibfnamefont
  {S.}~\bibnamefont {Liu}}, \bibinfo {author} {\bibfnamefont {K.}~\bibnamefont
  {Barmak}}, \bibinfo {author} {\bibfnamefont {K.}~\bibnamefont {Watanabe}},
  \bibinfo {author} {\bibfnamefont {T.}~\bibnamefont {Taniguchi}}, \bibinfo
  {author} {\bibfnamefont {A.~H.}\ \bibnamefont {MacDonald}}, \bibinfo {author}
  {\bibfnamefont {J.}~\bibnamefont {Shan}}, \ and\ \bibinfo {author}
  {\bibfnamefont {K.~F.}\ \bibnamefont {Mak}},\ }\bibfield  {title} {\enquote
  {\bibinfo {title} {Simulation of {Hubbard} model physics in {WSe2}/{WS2}
  moiré superlattices},}\ }\href {\doibase 10.1038/s41586-020-2085-3}
  {\bibfield  {journal} {\bibinfo  {journal} {Nature}\ }\textbf {\bibinfo
  {volume} {579}},\ \bibinfo {pages} {353} (\bibinfo {year}
  {2020})}\BibitemShut {NoStop}%
\bibitem [{\citenamefont {Regan}\ \emph {et~al.}(2020)\citenamefont {Regan},
  \citenamefont {Wang}, \citenamefont {Jin}, \citenamefont {Bakti~Utama},
  \citenamefont {Gao}, \citenamefont {Wei}, \citenamefont {Zhao}, \citenamefont
  {Zhao}, \citenamefont {Zhang}, \citenamefont {Yumigeta}, \citenamefont
  {Blei}, \citenamefont {Carlström}, \citenamefont {Watanabe}, \citenamefont
  {Taniguchi}, \citenamefont {Tongay}, \citenamefont {Crommie}, \citenamefont
  {Zettl},\ and\ \citenamefont {Wang}}]{regan_mott_2020}%
  \BibitemOpen
  \bibfield  {author} {\bibinfo {author} {\bibfnamefont {E.~C.}\ \bibnamefont
  {Regan}}, \bibinfo {author} {\bibfnamefont {D.}~\bibnamefont {Wang}},
  \bibinfo {author} {\bibfnamefont {C.}~\bibnamefont {Jin}}, \bibinfo {author}
  {\bibfnamefont {M.~I.}\ \bibnamefont {Bakti~Utama}}, \bibinfo {author}
  {\bibfnamefont {B.}~\bibnamefont {Gao}}, \bibinfo {author} {\bibfnamefont
  {X.}~\bibnamefont {Wei}}, \bibinfo {author} {\bibfnamefont {S.}~\bibnamefont
  {Zhao}}, \bibinfo {author} {\bibfnamefont {W.}~\bibnamefont {Zhao}}, \bibinfo
  {author} {\bibfnamefont {Z.}~\bibnamefont {Zhang}}, \bibinfo {author}
  {\bibfnamefont {K.}~\bibnamefont {Yumigeta}}, \bibinfo {author}
  {\bibfnamefont {M.}~\bibnamefont {Blei}}, \bibinfo {author} {\bibfnamefont
  {J.~D.}\ \bibnamefont {Carlström}}, \bibinfo {author} {\bibfnamefont
  {K.}~\bibnamefont {Watanabe}}, \bibinfo {author} {\bibfnamefont
  {T.}~\bibnamefont {Taniguchi}}, \bibinfo {author} {\bibfnamefont
  {S.}~\bibnamefont {Tongay}}, \bibinfo {author} {\bibfnamefont
  {M.}~\bibnamefont {Crommie}}, \bibinfo {author} {\bibfnamefont
  {A.}~\bibnamefont {Zettl}}, \ and\ \bibinfo {author} {\bibfnamefont
  {F.}~\bibnamefont {Wang}},\ }\bibfield  {title} {\enquote {\bibinfo {title}
  {Mott and generalized {Wigner} crystal states in {WSe2}/{WS2} moiré
  superlattices},}\ }\href {\doibase 10.1038/s41586-020-2092-4} {\bibfield
  {journal} {\bibinfo  {journal} {Nature}\ }\textbf {\bibinfo {volume} {579}},\
  \bibinfo {pages} {359} (\bibinfo {year} {2020})}\BibitemShut {NoStop}%
\bibitem [{\citenamefont {Xu}\ \emph {et~al.}(2020)\citenamefont {Xu},
  \citenamefont {Liu}, \citenamefont {Rhodes}, \citenamefont {Watanabe},
  \citenamefont {Taniguchi}, \citenamefont {Hone}, \citenamefont {Elser},
  \citenamefont {Mak},\ and\ \citenamefont {Shan}}]{xu_correlated_2020}%
  \BibitemOpen
  \bibfield  {author} {\bibinfo {author} {\bibfnamefont {Y.}~\bibnamefont
  {Xu}}, \bibinfo {author} {\bibfnamefont {S.}~\bibnamefont {Liu}}, \bibinfo
  {author} {\bibfnamefont {D.~A.}\ \bibnamefont {Rhodes}}, \bibinfo {author}
  {\bibfnamefont {K.}~\bibnamefont {Watanabe}}, \bibinfo {author}
  {\bibfnamefont {T.}~\bibnamefont {Taniguchi}}, \bibinfo {author}
  {\bibfnamefont {J.}~\bibnamefont {Hone}}, \bibinfo {author} {\bibfnamefont
  {V.}~\bibnamefont {Elser}}, \bibinfo {author} {\bibfnamefont {K.~F.}\
  \bibnamefont {Mak}}, \ and\ \bibinfo {author} {\bibfnamefont
  {J.}~\bibnamefont {Shan}},\ }\bibfield  {title} {\enquote {\bibinfo {title}
  {Correlated insulating states at fractional fillings of moir\'{e}
  superlattices},}\ }\href {\doibase 10.1038/s41586-020-2868-6} {\bibfield
  {journal} {\bibinfo  {journal} {Nature}\ }\textbf {\bibinfo {volume} {587}},\
  \bibinfo {pages} {214} (\bibinfo {year} {2020})}\BibitemShut {NoStop}%
\bibitem [{\citenamefont {Huang}\ \emph {et~al.}(2021)\citenamefont {Huang},
  \citenamefont {Wang}, \citenamefont {Miao}, \citenamefont {Wang},
  \citenamefont {Li}, \citenamefont {Lian}, \citenamefont {Taniguchi},
  \citenamefont {Watanabe}, \citenamefont {Okamoto}, \citenamefont {Xiao},
  \citenamefont {Shi},\ and\ \citenamefont {Cui}}]{huang_correlated_2021}%
  \BibitemOpen
  \bibfield  {author} {\bibinfo {author} {\bibfnamefont {X.}~\bibnamefont
  {Huang}}, \bibinfo {author} {\bibfnamefont {T.}~\bibnamefont {Wang}},
  \bibinfo {author} {\bibfnamefont {S.}~\bibnamefont {Miao}}, \bibinfo {author}
  {\bibfnamefont {C.}~\bibnamefont {Wang}}, \bibinfo {author} {\bibfnamefont
  {Z.}~\bibnamefont {Li}}, \bibinfo {author} {\bibfnamefont {Z.}~\bibnamefont
  {Lian}}, \bibinfo {author} {\bibfnamefont {T.}~\bibnamefont {Taniguchi}},
  \bibinfo {author} {\bibfnamefont {K.}~\bibnamefont {Watanabe}}, \bibinfo
  {author} {\bibfnamefont {S.}~\bibnamefont {Okamoto}}, \bibinfo {author}
  {\bibfnamefont {D.}~\bibnamefont {Xiao}}, \bibinfo {author} {\bibfnamefont
  {S.-F.}\ \bibnamefont {Shi}}, \ and\ \bibinfo {author} {\bibfnamefont
  {Y.-T.}\ \bibnamefont {Cui}},\ }\bibfield  {title} {\enquote {\bibinfo
  {title} {Correlated insulating states at fractional fillings of the
  {WS2}/{WSe2} moiré lattice},}\ }\href {\doibase 10.1038/s41567-021-01171-w}
  {\bibfield  {journal} {\bibinfo  {journal} {Nature Physics}\ }\textbf
  {\bibinfo {volume} {17}},\ \bibinfo {pages} {715} (\bibinfo {year}
  {2021})}\BibitemShut {NoStop}%
\bibitem [{\citenamefont {Li}\ \emph {et~al.}(2021{\natexlab{a}})\citenamefont
  {Li}, \citenamefont {Jiang}, \citenamefont {Li}, \citenamefont {Zhang},
  \citenamefont {Kang}, \citenamefont {Zhu}, \citenamefont {Watanabe},
  \citenamefont {Taniguchi}, \citenamefont {Chowdhury}, \citenamefont {Fu},
  \citenamefont {Shan},\ and\ \citenamefont {Mak}}]{li_continuous_2021}%
  \BibitemOpen
  \bibfield  {author} {\bibinfo {author} {\bibfnamefont {T.}~\bibnamefont
  {Li}}, \bibinfo {author} {\bibfnamefont {S.}~\bibnamefont {Jiang}}, \bibinfo
  {author} {\bibfnamefont {L.}~\bibnamefont {Li}}, \bibinfo {author}
  {\bibfnamefont {Y.}~\bibnamefont {Zhang}}, \bibinfo {author} {\bibfnamefont
  {K.}~\bibnamefont {Kang}}, \bibinfo {author} {\bibfnamefont {J.}~\bibnamefont
  {Zhu}}, \bibinfo {author} {\bibfnamefont {K.}~\bibnamefont {Watanabe}},
  \bibinfo {author} {\bibfnamefont {T.}~\bibnamefont {Taniguchi}}, \bibinfo
  {author} {\bibfnamefont {D.}~\bibnamefont {Chowdhury}}, \bibinfo {author}
  {\bibfnamefont {L.}~\bibnamefont {Fu}}, \bibinfo {author} {\bibfnamefont
  {J.}~\bibnamefont {Shan}}, \ and\ \bibinfo {author} {\bibfnamefont {K.~F.}\
  \bibnamefont {Mak}},\ }\bibfield  {title} {\enquote {\bibinfo {title}
  {Continuous {Mott} transition in semiconductor moir\'{e} superlattices},}\
  }\href {\doibase 10.1038/s41586-021-03853-0} {\bibfield  {journal} {\bibinfo
  {journal} {Nature}\ }\textbf {\bibinfo {volume} {597}},\ \bibinfo {pages}
  {350} (\bibinfo {year} {2021}{\natexlab{a}})}\BibitemShut {NoStop}%
\bibitem [{\citenamefont {Ghiotto}\ \emph {et~al.}(2021)\citenamefont
  {Ghiotto}, \citenamefont {Shih}, \citenamefont {Pereira}, \citenamefont
  {Rhodes}, \citenamefont {Kim}, \citenamefont {Zang}, \citenamefont {Millis},
  \citenamefont {Watanabe}, \citenamefont {Taniguchi}, \citenamefont {Hone},
  \citenamefont {Wang}, \citenamefont {Dean},\ and\ \citenamefont
  {Pasupathy}}]{ghiotto_quantum_2021}%
  \BibitemOpen
  \bibfield  {author} {\bibinfo {author} {\bibfnamefont {A.}~\bibnamefont
  {Ghiotto}}, \bibinfo {author} {\bibfnamefont {E.-M.}\ \bibnamefont {Shih}},
  \bibinfo {author} {\bibfnamefont {G.~S. S.~G.}\ \bibnamefont {Pereira}},
  \bibinfo {author} {\bibfnamefont {D.~A.}\ \bibnamefont {Rhodes}}, \bibinfo
  {author} {\bibfnamefont {B.}~\bibnamefont {Kim}}, \bibinfo {author}
  {\bibfnamefont {J.}~\bibnamefont {Zang}}, \bibinfo {author} {\bibfnamefont
  {A.~J.}\ \bibnamefont {Millis}}, \bibinfo {author} {\bibfnamefont
  {K.}~\bibnamefont {Watanabe}}, \bibinfo {author} {\bibfnamefont
  {T.}~\bibnamefont {Taniguchi}}, \bibinfo {author} {\bibfnamefont {J.~C.}\
  \bibnamefont {Hone}}, \bibinfo {author} {\bibfnamefont {L.}~\bibnamefont
  {Wang}}, \bibinfo {author} {\bibfnamefont {C.~R.}\ \bibnamefont {Dean}}, \
  and\ \bibinfo {author} {\bibfnamefont {A.~N.}\ \bibnamefont {Pasupathy}},\
  }\bibfield  {title} {\enquote {\bibinfo {title} {Quantum criticality in
  twisted transition metal dichalcogenides},}\ }\href {\doibase
  10.1038/s41586-021-03815-6} {\bibfield  {journal} {\bibinfo  {journal}
  {Nature}\ }\textbf {\bibinfo {volume} {597}},\ \bibinfo {pages} {345}
  (\bibinfo {year} {2021})}\BibitemShut {NoStop}%
\bibitem [{\citenamefont {Morales-Dur\'{a}n}\ \emph {et~al.}(2021)\citenamefont
  {Morales-Dur\'{a}n}, \citenamefont {MacDonald},\ and\ \citenamefont
  {Potasz}}]{morales-duran_metal-insulator_2021}%
  \BibitemOpen
  \bibfield  {author} {\bibinfo {author} {\bibfnamefont {N.}~\bibnamefont
  {Morales-Dur\'{a}n}}, \bibinfo {author} {\bibfnamefont {A.~H.}\ \bibnamefont
  {MacDonald}}, \ and\ \bibinfo {author} {\bibfnamefont {P.}~\bibnamefont
  {Potasz}},\ }\bibfield  {title} {\enquote {\bibinfo {title} {Metal-insulator
  transition in transition metal dichalcogenide heterobilayer moiré
  superlattices},}\ }\href {\doibase 10.1103/PhysRevB.103.L241110} {\bibfield
  {journal} {\bibinfo  {journal} {Physical Review B}\ }\textbf {\bibinfo
  {volume} {103}},\ \bibinfo {pages} {L241110} (\bibinfo {year}
  {2021})}\BibitemShut {NoStop}%
\bibitem [{\citenamefont {Imada}\ \emph {et~al.}(1998)\citenamefont {Imada},
  \citenamefont {Fujimori},\ and\ \citenamefont
  {Tokura}}]{imada_metal-insulator_1998}%
  \BibitemOpen
  \bibfield  {author} {\bibinfo {author} {\bibfnamefont {M.}~\bibnamefont
  {Imada}}, \bibinfo {author} {\bibfnamefont {A.}~\bibnamefont {Fujimori}}, \
  and\ \bibinfo {author} {\bibfnamefont {Y.}~\bibnamefont {Tokura}},\
  }\bibfield  {title} {\enquote {\bibinfo {title} {Metal-insulator
  transitions},}\ }\href {\doibase 10.1103/RevModPhys.70.1039} {\bibfield
  {journal} {\bibinfo  {journal} {Reviews of Modern Physics}\ }\textbf
  {\bibinfo {volume} {70}},\ \bibinfo {pages} {1039} (\bibinfo {year}
  {1998})}\BibitemShut {NoStop}%
\bibitem [{\citenamefont {Jamei}\ \emph {et~al.}(2005)\citenamefont {Jamei},
  \citenamefont {Kivelson},\ and\ \citenamefont
  {Spivak}}]{jamei_universal_2005}%
  \BibitemOpen
  \bibfield  {author} {\bibinfo {author} {\bibfnamefont {R.}~\bibnamefont
  {Jamei}}, \bibinfo {author} {\bibfnamefont {S.}~\bibnamefont {Kivelson}}, \
  and\ \bibinfo {author} {\bibfnamefont {B.}~\bibnamefont {Spivak}},\
  }\bibfield  {title} {\enquote {\bibinfo {title} {Universal {Aspects} of
  {Coulomb}-{Frustrated} {Phase} {Separation}},}\ }\href {\doibase
  10.1103/PhysRevLett.94.056805} {\bibfield  {journal} {\bibinfo  {journal}
  {Physical Review Letters}\ }\textbf {\bibinfo {volume} {94}},\ \bibinfo
  {pages} {056805} (\bibinfo {year} {2005})}\BibitemShut {NoStop}%
\bibitem [{\citenamefont {Camjayi}\ \emph {et~al.}(2008)\citenamefont
  {Camjayi}, \citenamefont {Haule}, \citenamefont {Dobrosavljevic},\ and\
  \citenamefont {Kotliar}}]{camjayi_coulomb_2008}%
  \BibitemOpen
  \bibfield  {author} {\bibinfo {author} {\bibfnamefont {A.}~\bibnamefont
  {Camjayi}}, \bibinfo {author} {\bibfnamefont {K.}~\bibnamefont {Haule}},
  \bibinfo {author} {\bibfnamefont {V.}~\bibnamefont {Dobrosavljevic}}, \ and\
  \bibinfo {author} {\bibfnamefont {G.}~\bibnamefont {Kotliar}},\ }\bibfield
  {title} {\enquote {\bibinfo {title} {Coulomb correlations and the
  {Wigner}–{Mott} transition},}\ }\href {\doibase 10.1038/nphys1106}
  {\bibfield  {journal} {\bibinfo  {journal} {Nature Physics}\ }\textbf
  {\bibinfo {volume} {4}},\ \bibinfo {pages} {932} (\bibinfo {year}
  {2008})}\BibitemShut {NoStop}%
\bibitem [{\citenamefont {Amaricci}\ \emph {et~al.}(2010)\citenamefont
  {Amaricci}, \citenamefont {Camjayi}, \citenamefont {Haule}, \citenamefont
  {Kotliar}, \citenamefont {Tanaskovic},\ and\ \citenamefont
  {Dobrosavljevic}}]{amaricci_extended_2010}%
  \BibitemOpen
  \bibfield  {author} {\bibinfo {author} {\bibfnamefont {A.}~\bibnamefont
  {Amaricci}}, \bibinfo {author} {\bibfnamefont {A.}~\bibnamefont {Camjayi}},
  \bibinfo {author} {\bibfnamefont {K.}~\bibnamefont {Haule}}, \bibinfo
  {author} {\bibfnamefont {G.}~\bibnamefont {Kotliar}}, \bibinfo {author}
  {\bibfnamefont {D.}~\bibnamefont {Tanaskovic}}, \ and\ \bibinfo {author}
  {\bibfnamefont {V.}~\bibnamefont {Dobrosavljevic}},\ }\bibfield  {title}
  {\enquote {\bibinfo {title} {Extended {Hubbard} model: {Charge} ordering and
  {Wigner}-{Mott} transition},}\ }\href {\doibase 10.1103/PhysRevB.82.155102}
  {\bibfield  {journal} {\bibinfo  {journal} {Physical Review B}\ }\textbf
  {\bibinfo {volume} {82}},\ \bibinfo {pages} {155102} (\bibinfo {year}
  {2010})}\BibitemShut {NoStop}%
\bibitem [{\citenamefont {Pan}\ and\ \citenamefont
  {Das~Sarma}(2021)}]{pan_interaction-driven_2021}%
  \BibitemOpen
  \bibfield  {author} {\bibinfo {author} {\bibfnamefont {H.}~\bibnamefont
  {Pan}}\ and\ \bibinfo {author} {\bibfnamefont {S.}~\bibnamefont
  {Das~Sarma}},\ }\bibfield  {title} {\enquote {\bibinfo {title}
  {Interaction-{Driven} {Filling}-{Induced} {Metal}-{Insulator} {Transitions}
  in {2D} {Moiré} {Lattices}},}\ }\href {\doibase
  10.1103/PhysRevLett.127.096802} {\bibfield  {journal} {\bibinfo  {journal}
  {Physical Review Letters}\ }\textbf {\bibinfo {volume} {127}},\ \bibinfo
  {pages} {096802} (\bibinfo {year} {2021})}\BibitemShut {NoStop}%
\bibitem [{\citenamefont {Zang}\ \emph {et~al.}(2021)\citenamefont {Zang},
  \citenamefont {Wang}, \citenamefont {Cano},\ and\ \citenamefont
  {Millis}}]{zang_hartree-fock_2021}%
  \BibitemOpen
  \bibfield  {author} {\bibinfo {author} {\bibfnamefont {J.}~\bibnamefont
  {Zang}}, \bibinfo {author} {\bibfnamefont {J.}~\bibnamefont {Wang}}, \bibinfo
  {author} {\bibfnamefont {J.}~\bibnamefont {Cano}}, \ and\ \bibinfo {author}
  {\bibfnamefont {A.~J.}\ \bibnamefont {Millis}},\ }\bibfield  {title}
  {\enquote {\bibinfo {title} {Hartree-{Fock} study of the moiré {Hubbard}
  model for twisted bilayer transition metal dichalcogenides},}\ }\href
  {\doibase 10.1103/PhysRevB.104.075150} {\bibfield  {journal} {\bibinfo
  {journal} {Physical Review B}\ }\textbf {\bibinfo {volume} {104}},\ \bibinfo
  {pages} {075150} (\bibinfo {year} {2021})}\BibitemShut {NoStop}%
\bibitem [{\citenamefont {Pan}\ and\ \citenamefont
  {Sarma}(2021)}]{pan_interaction_2021}%
  \BibitemOpen
  \bibfield  {author} {\bibinfo {author} {\bibfnamefont {H.}~\bibnamefont
  {Pan}}\ and\ \bibinfo {author} {\bibfnamefont {S.~D.}\ \bibnamefont
  {Sarma}},\ }\bibfield  {title} {\enquote {\bibinfo {title} {Interaction range
  and temperature dependence of symmetry breaking in strongly correlated {2D}
  moir{\textbackslash}'e {TMD} bilayers},}\ }\href
  {http://arxiv.org/abs/2110.11330} {\bibfield  {journal} {\bibinfo  {journal}
  {arXiv:2110.11330 [cond-mat]}\ } (\bibinfo {year} {2021})},\ \bibinfo {note}
  {arXiv: 2110.11330}\BibitemShut {NoStop}%
\bibitem [{\citenamefont {Lee}\ \emph {et~al.}(2006)\citenamefont {Lee},
  \citenamefont {Nagaosa},\ and\ \citenamefont {Wen}}]{lee_doping_2006}%
  \BibitemOpen
  \bibfield  {author} {\bibinfo {author} {\bibfnamefont {P.~A.}\ \bibnamefont
  {Lee}}, \bibinfo {author} {\bibfnamefont {N.}~\bibnamefont {Nagaosa}}, \ and\
  \bibinfo {author} {\bibfnamefont {X.-G.}\ \bibnamefont {Wen}},\ }\bibfield
  {title} {\enquote {\bibinfo {title} {Doping a {Mott} insulator: {Physics} of
  high-temperature superconductivity},}\ }\href {\doibase
  10.1103/RevModPhys.78.17} {\bibfield  {journal} {\bibinfo  {journal} {Reviews
  of Modern Physics}\ }\textbf {\bibinfo {volume} {78}},\ \bibinfo {pages} {17}
  (\bibinfo {year} {2006})}\BibitemShut {NoStop}%
\bibitem [{\citenamefont {Senthil}\ \emph
  {et~al.}(2004{\natexlab{a}})\citenamefont {Senthil}, \citenamefont
  {Vishwanath}, \citenamefont {Balents}, \citenamefont {Sachdev},\ and\
  \citenamefont {Fisher}}]{senthil2004deconfined}%
  \BibitemOpen
  \bibfield  {author} {\bibinfo {author} {\bibfnamefont {T.}~\bibnamefont
  {Senthil}}, \bibinfo {author} {\bibfnamefont {A.}~\bibnamefont {Vishwanath}},
  \bibinfo {author} {\bibfnamefont {L.}~\bibnamefont {Balents}}, \bibinfo
  {author} {\bibfnamefont {S.}~\bibnamefont {Sachdev}}, \ and\ \bibinfo
  {author} {\bibfnamefont {M.~P.}\ \bibnamefont {Fisher}},\ }\bibfield  {title}
  {\enquote {\bibinfo {title} {Deconfined quantum critical points},}\
  }\href@noop {} {\bibfield  {journal} {\bibinfo  {journal} {Science}\ }\textbf
  {\bibinfo {volume} {303}},\ \bibinfo {pages} {1490} (\bibinfo {year}
  {2004}{\natexlab{a}})}\BibitemShut {NoStop}%
\bibitem [{\citenamefont {Senthil}\ \emph
  {et~al.}(2004{\natexlab{b}})\citenamefont {Senthil}, \citenamefont {Balents},
  \citenamefont {Sachdev}, \citenamefont {Vishwanath},\ and\ \citenamefont
  {Fisher}}]{senthil2004quantum}%
  \BibitemOpen
  \bibfield  {author} {\bibinfo {author} {\bibfnamefont {T.}~\bibnamefont
  {Senthil}}, \bibinfo {author} {\bibfnamefont {L.}~\bibnamefont {Balents}},
  \bibinfo {author} {\bibfnamefont {S.}~\bibnamefont {Sachdev}}, \bibinfo
  {author} {\bibfnamefont {A.}~\bibnamefont {Vishwanath}}, \ and\ \bibinfo
  {author} {\bibfnamefont {M.~P.~A.}\ \bibnamefont {Fisher}},\ }\bibfield
  {title} {\enquote {\bibinfo {title} {Quantum criticality beyond the
  landau-ginzburg-wilson paradigm},}\ }\href@noop {} {\bibfield  {journal}
  {\bibinfo  {journal} {Physical Review B}\ }\textbf {\bibinfo {volume} {70}},\
  \bibinfo {pages} {144407} (\bibinfo {year} {2004}{\natexlab{b}})}\BibitemShut
  {NoStop}%
\bibitem [{\citenamefont {Burkov}\ and\ \citenamefont
  {Balents}(2005)}]{burkov_superfluid-insulator_2005}%
  \BibitemOpen
  \bibfield  {author} {\bibinfo {author} {\bibfnamefont {A.~A.}\ \bibnamefont
  {Burkov}}\ and\ \bibinfo {author} {\bibfnamefont {L.}~\bibnamefont
  {Balents}},\ }\bibfield  {title} {\enquote {\bibinfo {title}
  {Superfluid-insulator transitions on the triangular lattice},}\ }\href
  {\doibase 10.1103/PhysRevB.72.134502} {\bibfield  {journal} {\bibinfo
  {journal} {Physical Review B}\ }\textbf {\bibinfo {volume} {72}},\ \bibinfo
  {pages} {134502} (\bibinfo {year} {2005})}\BibitemShut {NoStop}%
\bibitem [{\citenamefont {Xu}\ \emph {et~al.}(2021)\citenamefont {Xu},
  \citenamefont {Luo}, \citenamefont {Jian},\ and\ \citenamefont
  {Xu}}]{xu_metal-insulator_2021}%
  \BibitemOpen
  \bibfield  {author} {\bibinfo {author} {\bibfnamefont {Y.}~\bibnamefont
  {Xu}}, \bibinfo {author} {\bibfnamefont {Z.-X.}\ \bibnamefont {Luo}},
  \bibinfo {author} {\bibfnamefont {C.-M.}\ \bibnamefont {Jian}}, \ and\
  \bibinfo {author} {\bibfnamefont {C.}~\bibnamefont {Xu}},\ }\bibfield
  {title} {\enquote {\bibinfo {title} {Metal-{Insulator} {Transition} with
  {Charge} {Fractionalization}},}\ }\href {http://arxiv.org/abs/2106.14910}
  {\bibfield  {journal} {\bibinfo  {journal} {arXiv:2106.14910 [cond-mat]}\ }
  (\bibinfo {year} {2021})},\ \bibinfo {note} {arXiv: 2106.14910}\BibitemShut
  {NoStop}%
\bibitem [{\citenamefont {Lannert}\ \emph {et~al.}(2001)\citenamefont
  {Lannert}, \citenamefont {Fisher},\ and\ \citenamefont
  {Senthil}}]{lannert_quantum_2001}%
  \BibitemOpen
  \bibfield  {author} {\bibinfo {author} {\bibfnamefont {C.}~\bibnamefont
  {Lannert}}, \bibinfo {author} {\bibfnamefont {M.~P.~A.}\ \bibnamefont
  {Fisher}}, \ and\ \bibinfo {author} {\bibfnamefont {T.}~\bibnamefont
  {Senthil}},\ }\bibfield  {title} {\enquote {\bibinfo {title} {Quantum
  confinement transition in a d-wave superconductor},}\ }\href {\doibase
  10.1103/PhysRevB.63.134510} {\bibfield  {journal} {\bibinfo  {journal}
  {Physical Review B}\ }\textbf {\bibinfo {volume} {63}},\ \bibinfo {pages}
  {134510} (\bibinfo {year} {2001})}\BibitemShut {NoStop}%
\bibitem [{\citenamefont {Li}\ \emph {et~al.}(2021{\natexlab{b}})\citenamefont
  {Li}, \citenamefont {Li}, \citenamefont {Regan}, \citenamefont {Wang},
  \citenamefont {Zhao}, \citenamefont {Kahn}, \citenamefont {Yumigeta},
  \citenamefont {Blei}, \citenamefont {Taniguchi}, \citenamefont {Watanabe},
  \citenamefont {Tongay}, \citenamefont {Zettl}, \citenamefont {Crommie},\ and\
  \citenamefont {Wang}}]{li_imaging_2021}%
  \BibitemOpen
  \bibfield  {author} {\bibinfo {author} {\bibfnamefont {H.}~\bibnamefont
  {Li}}, \bibinfo {author} {\bibfnamefont {S.}~\bibnamefont {Li}}, \bibinfo
  {author} {\bibfnamefont {E.~C.}\ \bibnamefont {Regan}}, \bibinfo {author}
  {\bibfnamefont {D.}~\bibnamefont {Wang}}, \bibinfo {author} {\bibfnamefont
  {W.}~\bibnamefont {Zhao}}, \bibinfo {author} {\bibfnamefont {S.}~\bibnamefont
  {Kahn}}, \bibinfo {author} {\bibfnamefont {K.}~\bibnamefont {Yumigeta}},
  \bibinfo {author} {\bibfnamefont {M.}~\bibnamefont {Blei}}, \bibinfo {author}
  {\bibfnamefont {T.}~\bibnamefont {Taniguchi}}, \bibinfo {author}
  {\bibfnamefont {K.}~\bibnamefont {Watanabe}}, \bibinfo {author}
  {\bibfnamefont {S.}~\bibnamefont {Tongay}}, \bibinfo {author} {\bibfnamefont
  {A.}~\bibnamefont {Zettl}}, \bibinfo {author} {\bibfnamefont {M.~F.}\
  \bibnamefont {Crommie}}, \ and\ \bibinfo {author} {\bibfnamefont
  {F.}~\bibnamefont {Wang}},\ }\bibfield  {title} {\enquote {\bibinfo {title}
  {Imaging two-dimensional generalized {Wigner} crystals},}\ }\href {\doibase
  10.1038/s41586-021-03874-9} {\bibfield  {journal} {\bibinfo  {journal}
  {Nature}\ }\textbf {\bibinfo {volume} {597}},\ \bibinfo {pages} {650}
  (\bibinfo {year} {2021}{\natexlab{b}})}\BibitemShut {NoStop}%
\bibitem [{\citenamefont {Kaul}\ and\ \citenamefont
  {Sachdev}(2008)}]{kaul_quantum_2008}%
  \BibitemOpen
  \bibfield  {author} {\bibinfo {author} {\bibfnamefont {R.~K.}\ \bibnamefont
  {Kaul}}\ and\ \bibinfo {author} {\bibfnamefont {S.}~\bibnamefont {Sachdev}},\
  }\bibfield  {title} {\enquote {\bibinfo {title} {Quantum criticality of
  {U}(1) gauge theories with fermionic and bosonic matter in two spatial
  dimensions},}\ }\href {\doibase 10.1103/PhysRevB.77.155105} {\bibfield
  {journal} {\bibinfo  {journal} {Physical Review B}\ }\textbf {\bibinfo
  {volume} {77}},\ \bibinfo {pages} {155105} (\bibinfo {year}
  {2008})}\BibitemShut {NoStop}%
\bibitem [{\citenamefont {Osborn}\ and\ \citenamefont
  {Stergiou}(2018)}]{osborn_seeking_2018}%
  \BibitemOpen
  \bibfield  {author} {\bibinfo {author} {\bibfnamefont {H.}~\bibnamefont
  {Osborn}}\ and\ \bibinfo {author} {\bibfnamefont {A.}~\bibnamefont
  {Stergiou}},\ }\bibfield  {title} {\enquote {\bibinfo {title} {Seeking fixed
  points in multiple coupling scalar theories in the $\epsilon$ expansion},}\
  }\href {\doibase 10.1007/JHEP05(2018)051} {\bibfield  {journal} {\bibinfo
  {journal} {Journal of High Energy Physics}\ }\textbf {\bibinfo {volume}
  {2018}},\ \bibinfo {pages} {51} (\bibinfo {year} {2018})}\BibitemShut
  {NoStop}%
\bibitem [{\citenamefont {Henriksson}\ and\ \citenamefont
  {Stergiou}(2021)}]{henriksson_perturbative_2021}%
  \BibitemOpen
  \bibfield  {author} {\bibinfo {author} {\bibfnamefont {J.}~\bibnamefont
  {Henriksson}}\ and\ \bibinfo {author} {\bibfnamefont {A.}~\bibnamefont
  {Stergiou}},\ }\bibfield  {title} {\enquote {\bibinfo {title} {Perturbative
  and nonperturbative studies of {CFTs} with {MN} global symmetry},}\ }\href
  {\doibase 10.21468/SciPostPhys.11.1.015} {\bibfield  {journal} {\bibinfo
  {journal} {SciPost Physics}\ }\textbf {\bibinfo {volume} {11}},\ \bibinfo
  {pages} {015} (\bibinfo {year} {2021})}\BibitemShut {NoStop}%
\bibitem [{\citenamefont {Moshe}\ and\ \citenamefont
  {Zinn-Justin}(2003)}]{moshe_quantum_2003}%
  \BibitemOpen
  \bibfield  {author} {\bibinfo {author} {\bibfnamefont {M.}~\bibnamefont
  {Moshe}}\ and\ \bibinfo {author} {\bibfnamefont {J.}~\bibnamefont
  {Zinn-Justin}},\ }\bibfield  {title} {\enquote {\bibinfo {title} {Quantum
  field theory in the large {N} limit: a review},}\ }\href {\doibase
  10.1016/S0370-1573(03)00263-1} {\bibfield  {journal} {\bibinfo  {journal}
  {Physics Reports}\ }\textbf {\bibinfo {volume} {385}},\ \bibinfo {pages} {69}
  (\bibinfo {year} {2003})}\BibitemShut {NoStop}%
\bibitem [{\citenamefont {Witczak-Krempa}\ \emph {et~al.}(2012)\citenamefont
  {Witczak-Krempa}, \citenamefont {Ghaemi}, \citenamefont {Senthil},\ and\
  \citenamefont {Kim}}]{witczak-krempa_universal_2012}%
  \BibitemOpen
  \bibfield  {author} {\bibinfo {author} {\bibfnamefont {W.}~\bibnamefont
  {Witczak-Krempa}}, \bibinfo {author} {\bibfnamefont {P.}~\bibnamefont
  {Ghaemi}}, \bibinfo {author} {\bibfnamefont {T.}~\bibnamefont {Senthil}}, \
  and\ \bibinfo {author} {\bibfnamefont {Y.~B.}\ \bibnamefont {Kim}},\
  }\bibfield  {title} {\enquote {\bibinfo {title} {Universal transport near a
  quantum critical {Mott} transition in two dimensions},}\ }\href {\doibase
  10.1103/PhysRevB.86.245102} {\bibfield  {journal} {\bibinfo  {journal}
  {Physical Review B}\ }\textbf {\bibinfo {volume} {86}},\ \bibinfo {pages}
  {245102} (\bibinfo {year} {2012})}\BibitemShut {NoStop}%
\bibitem [{\citenamefont {Zou}\ and\ \citenamefont
  {Chowdhury}(2020)}]{zou_deconfined_2020}%
  \BibitemOpen
  \bibfield  {author} {\bibinfo {author} {\bibfnamefont {L.}~\bibnamefont
  {Zou}}\ and\ \bibinfo {author} {\bibfnamefont {D.}~\bibnamefont
  {Chowdhury}},\ }\bibfield  {title} {\enquote {\bibinfo {title} {Deconfined
  metallic quantum criticality: {A} {U} ( 2 ) gauge-theoretic approach},}\
  }\href {\doibase 10.1103/PhysRevResearch.2.023344} {\bibfield  {journal}
  {\bibinfo  {journal} {Physical Review Research}\ }\textbf {\bibinfo {volume}
  {2}},\ \bibinfo {pages} {023344} (\bibinfo {year} {2020})}\BibitemShut
  {NoStop}%
\bibitem [{\citenamefont {Zhang}\ and\ \citenamefont {Sachdev}(2020)}]{YZSS20}%
  \BibitemOpen
  \bibfield  {author} {\bibinfo {author} {\bibfnamefont {Y.-H.}\ \bibnamefont
  {Zhang}}\ and\ \bibinfo {author} {\bibfnamefont {S.}~\bibnamefont
  {Sachdev}},\ }\bibfield  {title} {\enquote {\bibinfo {title} {{Deconfined
  criticality and ghost Fermi surfaces at the onset of antiferromagnetism in a
  metal}},}\ }\href {\doibase 10.1103/PhysRevB.102.155124} {\bibfield
  {journal} {\bibinfo  {journal} {Phys. Rev. B}\ }\textbf {\bibinfo {volume}
  {102}},\ \bibinfo {pages} {155124} (\bibinfo {year} {2020})}\BibitemShut
  {NoStop}%
\bibitem [{\citenamefont {Mandal}(2020)}]{mandal_critical_2020}%
  \BibitemOpen
  \bibfield  {author} {\bibinfo {author} {\bibfnamefont {I.}~\bibnamefont
  {Mandal}},\ }\bibfield  {title} {\enquote {\bibinfo {title} {Critical {Fermi}
  surfaces in generic dimensions arising from transverse gauge field
  interactions},}\ }\href {\doibase 10.1103/PhysRevResearch.2.043277}
  {\bibfield  {journal} {\bibinfo  {journal} {Physical Review Research}\
  }\textbf {\bibinfo {volume} {2}},\ \bibinfo {pages} {043277} (\bibinfo {year}
  {2020})}\BibitemShut {NoStop}%
\bibitem [{\citenamefont {Fisher}(1990)}]{fisher_quantum_1990}%
  \BibitemOpen
  \bibfield  {author} {\bibinfo {author} {\bibfnamefont {M.~P.~A.}\
  \bibnamefont {Fisher}},\ }\bibfield  {title} {\enquote {\bibinfo {title}
  {Quantum phase transitions in disordered two-dimensional superconductors},}\
  }\href {\doibase 10.1103/PhysRevLett.65.923} {\bibfield  {journal} {\bibinfo
  {journal} {Physical Review Letters}\ }\textbf {\bibinfo {volume} {65}},\
  \bibinfo {pages} {923} (\bibinfo {year} {1990})}\BibitemShut {NoStop}%
\bibitem [{\citenamefont {Ioffe}\ and\ \citenamefont
  {Larkin}(1989)}]{ioffe_gapless_1989}%
  \BibitemOpen
  \bibfield  {author} {\bibinfo {author} {\bibfnamefont {L.~B.}\ \bibnamefont
  {Ioffe}}\ and\ \bibinfo {author} {\bibfnamefont {A.~I.}\ \bibnamefont
  {Larkin}},\ }\bibfield  {title} {\enquote {\bibinfo {title} {Gapless fermions
  and gauge fields in dielectrics},}\ }\href {\doibase
  10.1103/PhysRevB.39.8988} {\bibfield  {journal} {\bibinfo  {journal}
  {Physical Review B}\ }\textbf {\bibinfo {volume} {39}},\ \bibinfo {pages}
  {8988} (\bibinfo {year} {1989})}\BibitemShut {NoStop}%
\end{thebibliography}%
\end{document}